%% file: main.tex
\documentclass[11pt]{article}
\usepackage[top=1in,left=1in,right=1in,bottom=1in]{geometry}
\pdfoutput=1

\newfont{\mycrnotice}{ptmr8t at 7pt}
\newfont{\myconfname}{ptmri8t at 7pt}
\let \originalleft \left
\let\originalright\right
\renewcommand{\left}{\mathopen{}\mathclose\bgroup\originalleft}
\renewcommand{\right}{\aftergroup\egroup\originalright}

\usepackage{enumitem}
\usepackage{booktabs}   %% For formal tables:
\usepackage{subcaption} %% For complex figures with subfigures/subcaptions
\usepackage{inputs/style}
\input{inputs/utils}

\usepackage{microtype}
\usepackage{bm}
\usepackage{color}          % Code coloring
\usepackage{zi4}            % Nice font
\usepackage[frozencache,cachedir=.]{minted}

\usepackage{xcolor}
\usepackage{mdframed}% http://ctan.org/pkg/mdframed

\definecolor[named]{ACMBlue}{cmyk}{1,0.1,0,0.1}
\definecolor[named]{ACMYellow}{cmyk}{0,0.16,1,0}
\definecolor[named]{ACMOrange}{cmyk}{0,0.42,1,0.01}
\definecolor[named]{ACMRed}{cmyk}{0,0.90,0.86,0}
\definecolor[named]{ACMLightBlue}{cmyk}{0.49,0.01,0,0}
\definecolor[named]{ACMGreen}{cmyk}{0.20,0,1,0.19}
\definecolor[named]{ACMPurple}{cmyk}{0.55,1,0,0.15}
\definecolor[named]{ACMDarkBlue}{cmyk}{1,0.58,0,0.21}
\usepackage[bookmarksnumbered,unicode]{hyperref}
\hypersetup{colorlinks,
  linkcolor=ACMRed,
  citecolor=ACMPurple,
  urlcolor=ACMDarkBlue,
  filecolor=ACMDarkBlue}

\begin{document}

\date{}

\title{Sage: Parallel Semi-Asymmetric Graph Algorithms for NVRAMs
\footnote{
This is an extended version of a paper in PVLDB (to be presented at VLDB'20).
The authors can be contacted at \{ldhulipa,
cmcguffe\}@cs.cmu.edu, kanghongbothu@gmail.com, ygu@cs.ucr.edu, \{guyb, gibbons\}@cs.cmu.edu
and jshun@mit.edu.}
}

\author{
  Laxman Dhulipala\\ CMU\and
  Charles McGuffey\\ CMU\and
  Hongbo Kang\\ Tsinghua\and
  Yan Gu\\ UC Riverside\and
  Guy E. Blelloch\\CMU\and
  Phillip B. Gibbons\\CMU\and
  Julian Shun\\MIT CSAIL}

\input{inputs/macro}

\maketitle

\input{inputs/abstract}
\input{inputs/intro}
\input{inputs/prelims}
\input{inputs/semiasymmetric}

\input{inputs/algorithms}

\input{inputs/experiments}

\input{inputs/related}
\input{inputs/conclusion}

\section*{Acknowledgements} This research was supported by DOE
Early Career Award DE-SC0018947, NSF CAREER Award CCF-1845763,
NSF grants CCF-1725663, CCF-1910030 and CCF-1919223, Google Faculty Research
Award, DARPA SDH Award HR0011-18-3-0007, and the Applications Driving
Algorithms (ADA) Center, a JUMP Center co-sponsored by SRC
and DARPA.

\bibliographystyle{abbrv}
\bibliography{new-ref}

\clearpage
\input{inputs/appendix}

\end{document}

%% file: inputs/utils.tex
\usepackage[most]{tcolorbox}
\tcbuselibrary{breakable}

\newtcolorbox{OuterBox}[1][]{%
    breakable,
    enhanced,
    frame hidden,
    interior hidden,
    left=-5pt,
    right=-5pt,
    top=-5pt,
    float=p,
    boxsep=0pt,
    arc=0pt
#1}%

\newtcolorbox{InnerBox}[1][]{%
    enforce breakable,
    enhanced,
    colback=gray,
    colframe=white,
#1}%

\newenvironment{tbox}{
\vspace{0.2cm}
\begin{tcolorbox}[width=\textwidth,
                  enhanced,
                  boxsep=1pt,
                  left=1pt,
                  right=1pt,
                  top=0.5pt,
                  boxrule=1pt,
                  arc=0pt,
                  colback=white,
                  colframe=black,
                  unbreakable
                  ]%%
}{
\end{tcolorbox}
}

\newenvironment{mytbox}{
\vspace{0.2cm}
\begin{tcolorbox}[width=\textwidth,
                  enhanced,
                  boxsep=1pt,
                  left=1pt,
                  right=1pt,
                  top=4pt,
                  boxrule=1pt,
                  arc=0pt,
                  colback=white,
                  colframe=black,
                  unbreakable
                  ]%%
}{
\end{tcolorbox}
}

\newcommand{\tboxhrule}[0]{\vspace{0.1cm} {\color{black} \hrule} \vspace{0.2cm}}

\newenvironment{mytitledtbox}[1]{\begin{mytbox}#1 \tboxhrule}{\end{mytbox}}
\newenvironment{benchmark}[1]{\begin{mytitledtbox}{{#1}}}{\end{mytitledtbox}}

%% file: inputs/macro.tex
\newcommand{\dist}{\ensuremath{\text{dist}}}

%% Graphs
\newcommand{\degree}[1]{\ensuremath{\textit{deg}(#1)}}
\newcommand{\diam}[1]{\ensuremath{\textit{diam}(#1)}}
\newcommand{\davg}{\ensuremath{d_{\textit{avg}}}}

%% Ligra
\newcommand{\vset}{vertexSubset}
\newcommand{\vsetaug}{vertexSubset}
\newcommand{\emap}{\textproc{edgeMap}}
\newcommand{\emapdata}{\textproc{edgeMapData}}
\newcommand{\ctngh}{\textproc{edgeMapCount}}
\newcommand{\emapfil}{\textproc{edgeMapFilter}}
\newcommand{\emapapp}{\textproc{edgeMapApply}}
\newcommand{\emapred}{\textproc{edgeMapReduce}}
\newcommand{\emapct}{\textproc{edgeMapSum}}
\newcommand{\vmap}{\textproc{vertexMap}}
\newcommand{\vfilter}{\textproc{vertexFilter}}
\newcommand{\emapdense}{\textproc{edgeMapDense}}
\newcommand{\emapsparse}{\textproc{edgeMapSparse}}
\newcommand{\emapblock}{\textproc{edgeMapBlocked}}
\newcommand{\notacq}{\textproc{notAcq}}
\newcommand{\acq}{\textproc{acq}}
\newcommand{\unacq}{\textproc{unacq}}
\newcommand{\emapdensewrite}{edgeMapDense-Write}
\newcommand{\cas}{CAS}
\newcommand{\ts}{test\_and\_set}
\newcommand{\fa}{fetch\_and\_add}
\newcommand{\writemin}{writeMin}
\newcommand{\update}{Update}
\newcommand{\cond}{Cond}
\newcommand{\maybe}{\textproc{maybe}}

%% Bucketing
\newcommand{\makebkt}{\textproc{makeBuckets}}
\newcommand{\nextbkt}{\textproc{nextBucket}}
\newcommand{\getbkt}{\textproc{getBucket}}
\newcommand{\updatebkt}{\textproc{updateBuckets}}
\newcommand{\bktorder}{\textproc{bucket\_order}}
\newcommand{\orderinc}{\textsc{increasing}}
\newcommand{\orderdec}{\textsc{decreasing}}
\newcommand{\prv}{\textsc{prev}}
\newcommand{\nxt}{\textsc{next}}
\newcommand{\curbkt}{\textsc{cur}}
\newcommand{\nullbkt}{\textsc{nullbkt}}
\newcommand{\size}{\textproc{size}}
\newcommand{\codevar}[1]{\mathit{#1}}
\newcommand{\mfloor}[1]{\left \lfloor #1 \right \rfloor }
\newcommand{\bkttype}[1]{\mathsf{#1}}
\newcommand{\ident}{\bkttype{identifier}}
\newcommand{\idents}{\bkttype{identifiers}}
\newcommand{\buckets}{\bkttype{buckets}}
\newcommand{\bktid}{\bkttype{bucket\_id}}
\newcommand{\bktids}{\bkttype{bucket\_ids}}
\newcommand{\bktdst}{\bkttype{bucket\_dest}}
\newcommand{\bktord}{\bkttype{bucket\_order}}

\newcommand{\src}{\textsc{src}}
\newcommand{\NC}{\mathsf{NC}}
\newcommand{\RNC}{\mathsf{RNC}}
\newcommand{\boruvka}{Bor\r{u}vka}
\newcommand{\Boruvka}{Bor\r{u}vka}

\newcommand{\tsmod}{$\mathsf{TS}$}
\newcommand{\pwmod}{$\mathsf{PW}$}
\newcommand{\famod}{$\mathsf{FA}$}
\newcommand{\ram}{$\mathsf{RAM}$}
\newcommand{\mpram}{$\mathsf{TRAM}$}
\newcommand{\pram}{$\mathsf{PRAM}$}
\newcommand{\crcwpram}{$\mathsf{CRCW}\ \mathsf{PRAM}$}
\newcommand{\tmpram}{BinaryForking-RAM}
\newcommand{\process}{thread}
\newcommand{\processes}{threads}

\newcommand{\nvm}{NVRAM}
\newcommand{\intelnvm}{Optane DC Persistent Memory}
\newcommand{\memorymode}{Memory Mode}
\newcommand{\appdirectmode}{App-Direct Mode}
\newcommand{\semiasym}{Semi-Asymmetric}
\newcommand{\semiextern}{semi-external}
\newcommand{\omegaind}{$\omega$-independent}
\newcommand{\SAM}{Parallel Semi-Asymmetric Model}
\newcommand{\SAMabbrev}{PSAM}
\newcommand{\BSAM}{Blocked-PSAM}
\newcommand{\SEM}{Semi-External Memory}
\newcommand{\EM}{External Memory}
\newcommand{\SSt}{semi-streaming}
\newcommand{\thread}{thread}
\newcommand{\wcost}{\omega}
\newcommand{\spanc}{span}

\newcommand{\largemem}{large-memory}
\newcommand{\smallmem}{small-memory}
\newcommand{\forkins}{\texttt{fork}}

\newcommand{\emapchunk}{\textproc{edgeMapChunked}}

%% Filtering
\newcommand{\gfilter}{graphFilter}
\newcommand{\asymgfilter}{asymGraphFilter}
\newcommand{\symgfilter}{symGraphFilter}
\newcommand{\makefilter}{\textproc{makeFilter}}
\newcommand{\makeasymfilter}{\textproc{makeAsymFilter}}
\newcommand{\filteredges}{\textproc{filterEdges}}
\newcommand{\vtxdegree}{\textproc{Degree}}
\newcommand{\emappack}{\textproc{edgeMapPack}}
\newcommand{\filterbs}{\ensuremath{\mathcal{F}_B}}
\newcommand{\wordsize}{\ensuremath{\mathcal{W}}}

\newcommand{\oursystem}{Sage}
\newcommand{\oursystemfull}{Semi-Asymmetric Graph Engine}

% revision
\definecolor{revise-color}{rgb}{0.76, 0.23, 0.13}
\definecolor{mygreen}{rgb}{0.0, 0.5, 0.0}
\definecolor{amaranth}{rgb}{0.9, 0.17, 0.31}
\newcommand{\revised}[1]{{#1}}

%% file: inputs/abstract.tex
\begin{abstract}
Non-volatile main memory (NVRAM) technologies provide an attractive
set of features for large-scale graph analytics, including
byte-addressability, low idle power, and improved memory-density.
\revised{NVRAM systems today have an order of magnitude more NVRAM than
traditional memory (DRAM). NVRAM systems could therefore potentially
allow very large graph problems to be solved on a single machine, at a
modest cost. However, a significant challenge in achieving high
performance is in accounting for the fact that NVRAM writes can be
much more expensive than NVRAM reads.}

In this paper, we propose an approach to parallel graph analytics
using the \emph{Parallel Semi-Asymmetric Model (PSAM)}, in which the
graph is stored as a read-only data structure (in NVRAM), and the
amount of mutable memory is kept proportional to the number of
vertices.  Similar to the popular semi-external and semi-streaming
models for graph analytics, the PSAM approach assumes that the
vertices of the graph fit in a fast read-write memory (DRAM), but the
edges do not. In NVRAM systems, our approach eliminates writes to the
NVRAM, among other benefits.

To experimentally study this new setting, we develop
\emph{\oursystem{}}, a parallel semi-asymmetric graph engine with
which we implement provably-efficient (and often work-optimal)
PSAM algorithms for over a dozen fundamental graph problems.
We experimentally study \oursystem{} using a 48-core
machine on the largest publicly-available real-world graph (the
Hyperlink Web graph with over 3.5 billion vertices and 128 billion
edges) equipped with \intelnvm{}, and show that \oursystem{}
outperforms the fastest prior systems designed for NVRAM. Importantly,
we also show that \oursystem{} nearly matches the fastest prior
systems running solely in DRAM, by effectively hiding the costs of
repeatedly accessing NVRAM versus DRAM.
\end{abstract}

%% file: inputs/intro.tex
\section{Introduction}\label{sec:intro}

Over the past decade, there has been a steady increase in the
main-memory sizes of commodity multicore machines, which has led to
the development of fast single-machine shared-memory graph algorithms
for processing massive graphs with hundreds of billions of
edges~\cite{dhulipala18scalable,nguyen2013lightweight,ShunB2013,shun2015ligraplus}
on a single machine. Single-machine analytics by-and-large outperform
their distributed memory counterparts, running up to \emph{orders of
magnitude faster} using much fewer
resources~\cite{dhulipala18scalable,mcsherry2015cost,ShunB2013,
shun2015ligraplus}. These analytics have become
increasingly relevant due to a longterm trend of increasing memory
sizes, which continues today in the form of new non-volatile memory
technologies that are now emerging on the market (e.g., Intel's
\emph{Optane DC Persistent Memory}). These devices are significantly cheaper
on a per-gigabyte basis, provide an order of magnitude greater memory
capacity per DIMM than traditional DRAM, and offer byte-addressability
and low idle power, thereby providing a realistic and cost-efficient
way to equip a commodity multicore machine with multiple terabytes of
non-volatile RAM (NVRAM).

\input{inputs/figure_us_vs_memmode}

Due to these advantages, NVRAMs are likely to be a key component of
many future memory hierarchies, likely in conjunction with a smaller
amount of traditional DRAM.  However, a challenge of these
technologies is to overcome an \emph{asymmetry} between reads and
writes---write operations are more expensive than reads in terms of
energy and throughput.  This property requires rethinking algorithm
design and implementations to minimize the number of writes to
NVRAM~\cite{BBFGGMS16, BFGGS15, blelloch2016efficient,
carson2016write, viglas2014write}.  As an
example of the memory technology and its tradeoffs, in this paper we
use a 48-core machine that has 8x as much NVRAM as
DRAM (we are aware of machines with 16x as much NVRAM as
DRAM~\cite{Gill2019}), where the combined read throughput for all cores
from the NVRAM is about 3x slower than reads from the DRAM, and writes
on the NVRAM are a further factor of about 4x
slower~\cite{izraelevitz2019basic, van2019persistent} (a factor of 12
total). Under this asymmetric setting, algorithms performing a large
number of writes could see a significant performance penalty if care
is not taken to avoid or eliminate writes.

An important property of most graphs used in practice is that they are
sparse, but still tend to have many more edges than vertices, often
from one to two orders of magnitude more.  This is true for almost all
social network graphs~\cite{leskovec2014snap}, but also for many
graphs that are derived from various simulations~\cite{SuiteSparse}.
In Figure~\ref{fig:avg_degree} we show that over 90\% of the large
graphs (more than 1 million vertices) from the
SNAP~\cite{leskovec2014snap} and LAW~\cite{boldi2004webgraph}
datasets have at least 10 times as many edges as vertices.  Given
that very large graphs today can have over 100 billion edges
(requiring around a terabyte of storage), but only a few billion
vertices, a popular and reasonable assumption both in theory and in
practice is that vertices, but not edges, fit in
DRAM~\cite{Abello2002, feigenbaum2005graph, Kliemann16, McGregor14,
Mehlhorn02, Muthukrishnan2005, pearce2010multithreaded, SunWDX17,
zheng15flashgraph, Zheng17}.

\input{inputs/figure_avg_degree}

With these characteristics of NVRAM and real-world graphs in mind, we
propose a \emph{semi-asymmetric} approach to parallel graph analytics,
in which (i) the full graph is stored in NVRAM and is accessed in
\emph{read-only mode} and (ii) the amount of DRAM is proportional to
the number of vertices.  Although completely avoiding writes to the
NVRAM may seem overly restrictive, the approach has the following
benefits: (i) algorithms avoid the high cost of NVRAM writes, (ii) the
algorithms do not contribute to NVRAM wear-out or wear-leveling
overheads, and (iii) algorithm design is independent of the actual
cost of NVRAM writes, which has been shown to vary based on access
pattern and number of
cores~\cite{izraelevitz2019basic,van2019persistent} and will likely
change with innovations in NVRAM technology and controllers.
Moreover, it enables an important NUMA optimization in which a copy of
the graph is stored on each socket (Section~\ref{sec:exps}), for fast
read-only access without any cross-socket coordination.  Finally, with
no graph mutations, there is no need to re-compress the graph
on-the-fly when processing compressed
graphs~\cite{dhulipala2017julienne, dhulipala18scalable}.

The key question, then, is the following:
\begin{center}
\emph{Is the (restrictive)
semi-asymmetric approach effective for designing fast graph
algorithms?}
\end{center}
In this paper, we provide strong theoretical and
experimental evidence of the approach's effectiveness.

Our main contribution is \emph{\oursystem{}}, a parallel
semi-asymmetric graph engine with which we implement
provably-efficient (and often work-optimal) algorithms for over a
dozen fundamental graph problems (see Table~\ref{table:introtable}).
The key innovations are in ensuring that the updated state is
associated with vertices and not edges, which is particularly
challenging (i) for certain edge-based parallel graph traversals and
(ii) for algorithms that ``delete'' edges as they go along in order to
avoid revisiting them once they are no longer needed.  We provide
general techniques (Sections~\ref{subsec:blocking}
and~\ref{sec:filter}) to solve these two problems.  For the
latter, used by four of our algorithms, we require relaxing the
prescribed amount of DRAM to be on the order of one bit per edge.
Details of our algorithms are given in
Section~\ref{sec:algorithms}. Our codes extend the current
state-of-the-art DRAM-only codes from GBBS~\cite{dhulipala18scalable},
% and are publicly available as part of GBBS,
and can be found at
\url{https://github.com/ParAlg/gbbs/tree/master/sage}.

From a theoretical perspective, we propose a model for analyzing
algorithms in the semi-asymmetric setting
(Section~\ref{sec:semiasymmetric}).  The model, called the \SAM{}
(\SAMabbrev), consists of a shared asymmetric \largemem{} with
unbounded size that can hold the entire graph, and a shared symmetric
\smallmem{} with $O(n)$ words of memory, where $n$ is the number of
vertices in the graph.  In a relaxed version of the model, we allow
\smallmem{} size of $O(n+m/\log n)$ words, where $m$ is the number of
edges in the graph. Although we do not use writes to the \largemem{}
in our algorithms, the \SAMabbrev{} model permits writes to the
\largemem{}, which are $\omega > 1$ times more costly than reads.
We prove strong theoretical bounds in terms of \SAMabbrev{} work and
depth for all of our parallel algorithms in \oursystem{}, as shown in
Table~\ref{table:introtable}. Most of the algorithms are
work-efficient (performing asymptotically the same work as the best
sequential algorithm for the problem) and have polylogarithmic depth
(parallel time). These provable guarantees ensure that our algorithms
perform reasonably well across graphs with different characteristics,
machines with different core counts, and \nvm{}s with different
read-write asymmetries.

We experimentally study \oursystem{} on large-scale real-world graphs
(Section~\ref{sec:exps}). We show that \oursystem{} scales to the
largest publicly-available graph, the Hyperlink2012 graph with over
3.5 billion vertices and 128 billion edges (and 225 billion edges for
algorithms running on the undirected/symmetrized graph).
Figure~\ref{fig:intro_us_vs_memmode} compares the performance of
\oursystem{} algorithms with the fastest available NVRAM approaches on
the Hyperlink2012 graph, which is larger than the DRAM of the
machine used in our experiments. Compared with the state-of-the-art
DRAM codes from GBBS~\cite{dhulipala18scalable}, automatically
extended to use NVRAM using MemoryMode,\footnote{Effectively using the DRAM as
a cache---see Section~\ref{sec:configurations}.} \oursystem{} is 1.89x faster
on average, and slower only in one instance for reasons that we
discuss in Section~\ref{sec:exps}. Compared with the recently
developed codes from~\cite{Gill2019}, which are the current
state-of-the-art NVRAM codes available today, our codes are faster
on all five graph problems studied in~\cite{Gill2019},
and achieve an average speedup of 1.94x.

We also study the performance of \oursystem{} compared with
state-of-the-art graph codes run \emph{entirely in DRAM} on the
largest dataset used in our study that still fits within memory
(Figure~\ref{fig:intro_us_vs_inmem} in
Section~\ref{subsec:usvsinmem}).
We compare \oursystem{} with the GBBS codes run entirely in DRAM, and
also when automatically converted to use NVRAM using libvmmalloc, a
standard NVRAM memory allocator. Compared with GBBS codes running in
DRAM, \oursystem{} on NVRAM is only 1.01x slower on average, within 6\% of the
in-memory running time on all but two problems, and at most 1.82x
slower in the worst case.  Interestingly, we find that \oursystem{} when run in DRAM
is 1.17x faster than the GBBS codes run in DRAM on average.  This
indicates that our optimizations to reduce writes also help on DRAM,
although not to the same extent as on NVRAM.
In particular,
\oursystem{} on NVRAM is 6.69x faster on average than GBBS when run on NVRAM
using libvmmalloc. Thus, \oursystem{} significantly outperforms a
naive approach to convert DRAM codes to NVRAM ones, is faster than
state-of-the-art DRAM-only codes when run in DRAM, and is highly
competitive with the fastest DRAM-only running times when run in
NVRAM.

We summarize our contributions below.
\begin{enumerate}[label=(\textbf{\arabic*}),topsep=1pt,itemsep=0pt,parsep=0pt,leftmargin=15pt]
\item A semi-asymmetric approach to parallel graph analytics that
avoids writing to the NVRAM and uses DRAM proportional to the number
of vertices.

\item \oursystem{}:~a parallel semi-asymmetric graph engine with
implementations of 18 fundamental graph problems, and general
techniques for semi-asymmetric parallel graph algorithms. We have made
all of our codes publicly-available.\footnote{Our code can be found at
\url{https://github.com/ParAlg/gbbs/tree/master/sage}, and an accompanying website
at \url{https://paralg.github.io/gbbs/sage}.}

\item A new theoretical model called the \SAM{}, and techniques for
designing efficient, and often work-optimal parallel graph algorithms
in the model.

\item A comprehensive experimental evaluation of \oursystem{} on an
NVRAM system showing that \oursystem{} significantly outperforms prior
work and nearly matches state-of-the-art DRAM-only performance.
\end{enumerate}

%% file: inputs/figure_us_vs_memmode.tex
\begin{figure*}[!t]
\centering
\hspace*{-0.6cm}
\includegraphics[width=\textwidth]{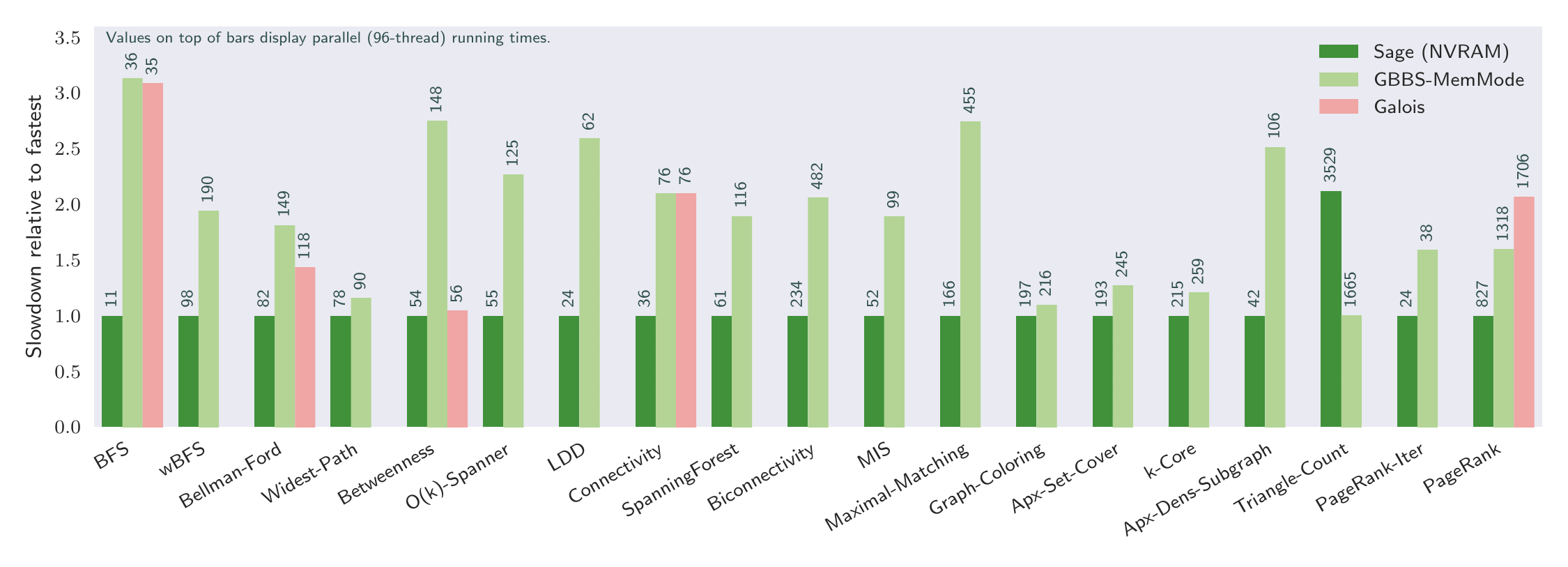}
%\vspace{-1em}
\caption{\label{fig:intro_us_vs_memmode}
Performance of \oursystem{} on the Hyperlink2012 graph compared with
existing state-of-the-art systems for processing
\emph{larger-than-memory graphs} using NVRAM measured relative to the
fastest system (smaller is better).  \oursystem{} (NVRAM) are the
new codes developed in this paper, GBBS-MemMode is the code
developed in \cite{dhulipala18scalable} run using MemoryMode,
and {Galois} is the NVRAM codes from
\cite{Gill2019}.  The bars are annotated with the parallel
running times (in seconds) of the codes on a 48-core system with 2-way
hyper-threading. Note that the Hyperlink2012 graph \emph{does not fit
in DRAM} for the machine used in these experiments.
}
\end{figure*}

%% file: inputs/figure_avg_degree.tex
\begin{figure}[!t]
\begin{center}
\vspace{-0.2em}
\includegraphics[scale=0.6]{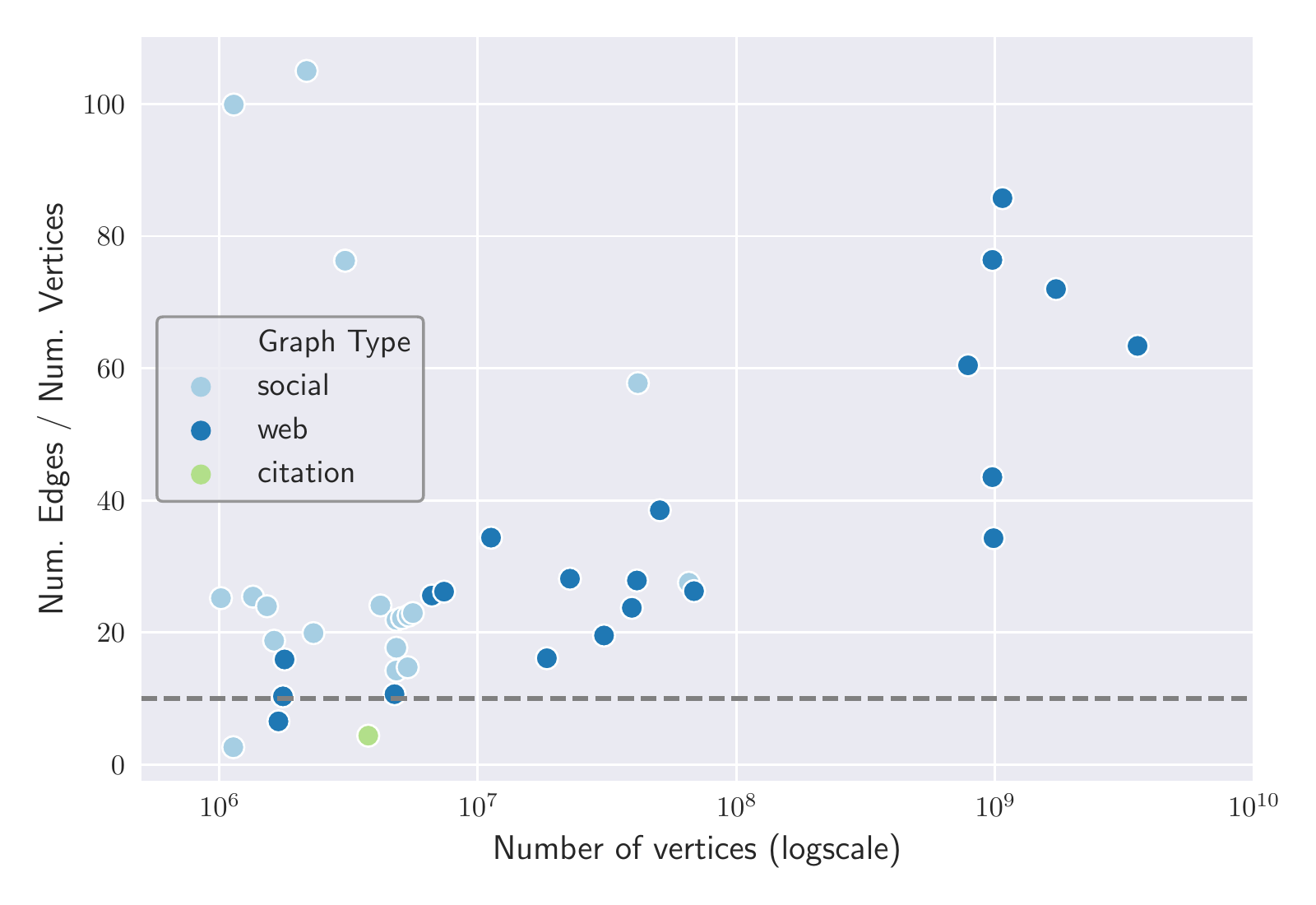}
\vspace{-0.6em}
\caption{\label{fig:avg_degree}
Number of vertices (logscale) vs. average degree $(m/n)$ on 42
real-world graphs with $n > 10^6$ from the
SNAP~\cite{leskovec2014snap} and LAW~\cite{boldi2004webgraph} datasets.
Over 90\% of the graphs have average degree larger than 10
(corresponding to the gray dashed line).
}
\end{center}
\vspace{-1em}
\end{figure}

%% file: inputs/prelims.tex
\section{Preliminaries}\label{sec:prelims}

\myparagraph{Graph Notation} We denote an unweighted graph by $G(V,
E)$, where $V$ is the set of vertices and $E$ is the set of edges.
The
number of vertices is $n = |V|$ and the number of edges is
$m = |E|$. Vertices are assumed to be indexed from $0$ to $n-1$. We
use $N(v)$ to denote the neighbors of vertex $v$ and \degree{v}
to denote its degree.
We focus on undirected graphs in this paper, although many of our
algorithms and techniques naturally generalize to directed graphs.  We
assume that $m = \Omega(n)$ when reporting bounds.  We use
$\mathsf{d}_{G}$ to refer to the diameter of the graph, which is the
longest shortest path distance between any vertex $s$ and any vertex
$v$ reachable from $s$. $\Delta$ ($\davg$) is used to denote the
maximum (average) degree of the graph. We assume that there are no
self-edges or duplicate edges in the graph.
We refer to graphs stored in the compressed sparse column and
compressed sparse row formats as \defn{CSC} and \defn{CSR},
respectively.  We also consider compressed graphs that store the
differences between consecutive neighbors using variable-length codes
for each sorted adjacency list~\cite{shun2015ligraplus}.

\myparagraph{Parallel Cost Model}
We use the work-depth model in this paper, and define the model
formally when introducing the \SAM{} model, which extends it
(Section~\ref{sec:semiasymmetric}).

\myparagraph{Parallel Primitives}
The following parallel procedures are used throughout the paper.
\defn{Prefix Sum (Scan)} takes as input an array $A$ of length $n$, an
associative binary operator $\oplus$, and an identity element $\bot$
such that $\bot \oplus x = x$ for any $x$, and returns the array
$(\bot, \bot \oplus A[0], \bot \oplus A[0] \oplus A[1], \ldots, \bot
\oplus_{i=0}^{n-2} A[i])$ as well as the overall sum, $\bot
\oplus_{i=0}^{n-1} A[i]$.  \defn{Reduce} takes an array $A$ and a
binary associative function $f$ and returns the sum of the elements in
$A$ with respect to $f$.  \defn{Filter} takes an array $A$ and a
predicate $f$ and returns a new array containing $a \in A$ for which
$f(a)$ is true, in the same order as in $A$.  If done in \smallmem{} (Section~\ref{sec:semiasymmetric}),
prefix sum, reduce and filter can all be done in $O(n)$ work and
$O(\log n)$ depth (assuming that $\oplus$ and $f$ take $O(1)$
work)~\cite{JaJa92}.

\myparagraph{Ligra, Ligra+, and Julienne}
As discussed in Section~\ref{sec:sage},
\oursystem{} builds on the Ligra~\cite{ShunB2013},
Ligra+~\cite{shun2015ligraplus}, and
Julienne~\cite{dhulipala2017julienne} frameworks for
shared-memory graph processing.
These frameworks provide primitives for representing subsets of vertices
(\defn{\vset{}}), and mapping functions over them (\defn{\emap{}}).
\emap{} takes as input a graph
$G(V,E)$, a \vset{} $U \subseteq V$, and two boolean functions $F$ and $C$.
\emap{} applies $F$ to $(u,v) \in E$ such that $u \in U$ and $C(v) =
\codevar{true}$ (call this subset of undirected edges $E_{a}$), and returns a
\vset{} $U'$ where $u \in U'$ if and only if $(u,v) \in E_{a}$ and
$F(u,v) = \codevar{true}$.  $F$ can side-effect data structures
associated with the vertices.

%% file: inputs/semiasymmetric.tex
\section{\SAM{}}\label{sec:semiasymmetric}

\subsection{Model Definition}

\input{inputs/figure-model}
The \defn{\SAM{} (\SAMabbrev)} consists of an \emph{asymmetric} \largemem{}
(NVRAM) with unbounded size, and a \emph{symmetric} \smallmem{} (DRAM)
with $O(n)$ words of memory. In a relaxed version of the model, we
allow \smallmem{} size of $O(n+m/\log n)$ words. The relaxed version
is intended to model a system where the ratio of \nvm{} to DRAM is
close to the average degree of real-world graphs (see
Figure~\ref{fig:avg_degree} and Table~\ref{table:graphinfo}).

The \SAMabbrev{} has a set of \thread{}s that share both the \largemem{} and
\smallmem{}. The underlying mechanisms for parallelism are identical
to the T-RAM or binary forking model, which is discussed in detail
in~\cite{blelloch18notes, blelloch2019optimal, dhulipala18scalable}.
In the model, each \thread{} acts like a sequential RAM that also
has a \forkins{} instruction. When a \thread{} performs a
\forkins{}, two newly created child \thread{}s run starting at the
next instruction, and the original \thread{} is suspended until all
the children terminate. A computation starts with a single
\emph{root} \thread{} and finishes when that root \thread{} finishes.

\myparagraph{Algorithm Cost}
We analyze algorithms on the \SAMabbrev{} using the \defn{work-depth measure}~\cite{JaJa92}.
The work-depth measure is a fundamental tool in analyzing parallel
algorithms, e.g., see ~\cite{BlellochFS16, dhulipala18scalable,
DBLP:conf/spaa/GuSSB15, shun2014practical, sun2019supporting, pam,
wang2019pardbscan} for a sample of recent practical
uses of this model.
Like other multi-level models (e.g., the ANP model~\cite{BBFGGMS16}),
we assume unit cost reads and writes to the \smallmem{}, and reads
from the \largemem{}, all in the unit of a word. A write to the
\largemem{} has a cost of $\wcost{}>1$, which is the cost of a write
relative to a read on NVRAMs.  The overall \defn{work} $W$ of an
algorithm is the sum of the costs for all memory accesses by all
\thread{}s.  The \defn{depth} $D$ is the cost of the highest cost
sequence of dependent instructions in the computation.  A
work-stealing scheduler can execute a computation in $W/p+O(D)$ time
with high probability on $p$ processors~\cite{BBFGGMS16, blumofe1999scheduling}.
Figure~\ref{fig:psam} illustrates the \SAMabbrev{} model.

\subsection{Discussion}\label{sec:psamdiscussion}
It is helpful to first clarify why we chose to keep the modeling
parameters simple, focusing on \emph{NVRAM read-write asymmetry}, when several
other parameters are also available (as discussed below).
Our goal was to design a theoretical model that helps guide algorithm
design by capturing the most salient features of the new hardware.

\myparagraph{Modeling Read and Write Costs}
Although NVRAM reads are about 3x more costly than accesses to
DRAM~\cite{van2019persistent}, we charge both unit cost in the
\SAMabbrev{}. When this cost gap needs to be studied (especially for
showing lower bounds), we can use an approach similar to the
asymmetric RAM (ARAM) model~\cite{blelloch2016efficient}, and define
the \emph{I/O cost $Q$} of an algorithm without charging for
instructions or DRAM accesses.  All algorithms in this paper have
asymptotically as many instructions as \nvm{} reads, and therefore
have the same I/O cost $Q$ as work $W$ up to constant factors.

\myparagraph{Writes to Large-memory}
Although in the approach used in this paper we do not perform writes
to the \largemem{}, the PSAM is designed to allow for analyzing
alternate approaches that do perform writes to \largemem{}.
Furthermore, permitting writes to the \largemem{} enables us to
consider the cost of algorithms from previous work such as
GBBS~\cite{dhulipala18scalable} and observe that many prior algorithms with $W$ work in the standard work-depth model are $\Theta(\omega
W)$ work in the PSAM.
We emphasize that the objective of this work is to
evaluate whether the restrictive approach used in our algorithms---i.e., completely avoiding writes to the \largemem, thereby gaining the benefits discussed in Section~\ref{sec:intro}---is effective compared to
existing approaches for programming NVRAM graph algorithms.
We note that algorithms designed with
a small number of \largemem{} writes could possibly be quite efficient in practice.

\myparagraph{Applicability}
In this paper, we provide evidence that the \SAMabbrev{} is broadly
applicable for many (18) fundamental graph problems.  We believe that many
other problems will also fit in the \SAMabbrev{}. For example,
counting and enumerating $k$-cliques, which were very recently studied
in the in-memory setting~\cite{shi2020parallel}, can be adapted to the
\SAMabbrev{} using the filtering technique proposed in this paper.
Other fundamental subgraph problems, such as subgraph
matching~\cite{han2013turboiso, sun2012efficient} and frequent
subgraph mining~\cite{elseidy2014grami, wang2018rstream} could be
solved in the \SAMabbrev{} using a similar approach, but mining
many large subgraphs may require performing some writes to the NVRAM (as
discussed below). Other problems, such as local search problems
including CoSimRank~\cite{rothe2014cosimrank}, personalized PageRank,
and other local clustering problems~\cite{shun2016parallel}, naturally
fit in the regular \SAMabbrev{} model.

We note that certain problems seem to require performing writes in the
\SAMabbrev{}. For example, in the $k$-truss problem, the output
requires emitting the trussness value for each edge, and thus storing
the output requires $\Theta(m)$ words of memory, which requires $\Theta(\omega
m)$ cost due to writes. Generalizations of $k$-truss, such as the $(r,s)$-nucleii
problem appear to have the same requirement for $r \geq
2$~\cite{sariyuce2015finding}.

\subsection{Relationship to Other Models}\label{sec:psamcomparison}
\vspace{-.2em}
\myparagraph{Asymmetric Models}
The model considered in this paper is related to the ARAM
model~\cite{blelloch2016efficient} and the asymmetric nested-parallel
(ANP) model~\cite{BBFGGMS16}.
Compared to these more general models, the PSAM is specially designed for graphs, with its \smallmem{} being either $O(n)$ or $O(n + m/\log n)$ words (for $n$ vertices and $m$ edges).

\myparagraph{External and Semi-External Memory Models}
The \EM{} model (also known as the I/O or disk-access
model)~\cite{AggarwalV88} is a classic two-level memory model
containing a bounded internal memory of size $M$ and an unbounded
external memory. I/Os to the external memory are done in blocks of
size $B$.
The \SEM{} model~\cite{Abello2002} is a relaxation
of the \EM{} model where there is a \smallmem{} that can hold
the vertices but not the edges.

There are three major differences between the \SAMabbrev{} and the
\EM{} and \SEM{} models. First, unlike the PSAM, neither the \EM{} nor the
\SEM{} model \emph{account for accessing the \smallmem{} (DRAM)},
because the objective of these models is to focus on the cost of
expensive I/Os to the external memory.  We believe that for existing
systems with NVRAMs, the cost of DRAM accesses is not negligible.
Second, both the \EM{} and \SEM{} have a parameter $B$ to model data
movement in large chunks.
NVRAMs support random access, so for the ease of design and analysis
we omit this parameter $B$.
Third, the \SAMabbrev{} explicitly models the asymmetry of writing
to the large memory, whereas the \EM{} and \SEM{} models treat both
reads and writes to the external memory indistinguishably (both cost
$B$). The asymmetry of these devices is significant for current
devices (writes to NVRAM are 4x slower than reads from NVRAM, and 12x
slower than reads from DRAM~\cite{izraelevitz2019basic,
van2019persistent}), and could be even larger in future generations of
energy-efficient NVRAMs. Explicitly modeling asymmetry is an important
aspect of our approach in the \SAMabbrev{}.

\myparagraph{Semi-Streaming Model}
In the \SSt{} model~\cite{feigenbaum2005graph, Muthukrishnan2005}, there is a memory size
of $O(n \cdot \text{polylog}(n))$ bits and algorithms can only read
the graph in a sequential streaming order (with possibly multiple
passes). In contrast, the \SAMabbrev{} allows random access to the input
graph because NVRAMs intrinsically support random access. Furthermore
the \SAMabbrev{} allows expensive writes to the \largemem{}, which is read-only in the \SSt{} model.

%% file: inputs/figure-model.tex
\begin{figure}
\begin{center}
\includegraphics[width=0.45\columnwidth]{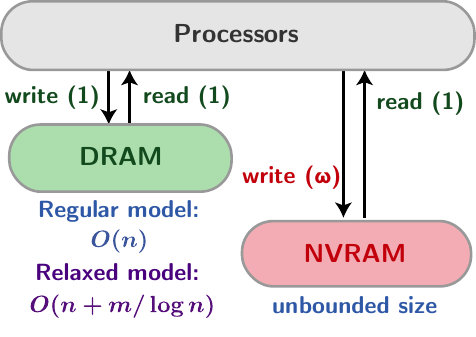}
\caption{\label{fig:psam}
The Parallel Semi-Asymmetric Model. Algorithms in the model perform
accesses to a symmetric \smallmem{} (DRAM) and an asymmetric \largemem{} (NVRAM)
at a word-granularity. Reads from both memories are charged unit-cost,
whereas writes to the asymmetric memory are charged $\bm{\omega}$. In the
regular model, algorithms have access to $O(n)$ words of symmetric
memory, and in a relaxed variant have access to $O(n + m / \log n)$ words
of symmetric memory. Compared to existing two-level models, the main
advantages of the \SAMabbrev{} are that it explicitly models
\emph{NVRAM read-write asymmetry} and it provides \emph{sufficient symmetric memory}
to design provably-efficient and practical parallel graph algorithms.
}
\vspace{-1em}
\end{center}
\end{figure}

%% file: inputs/algorithms.tex
\section{\oursystem{}: A Semi-Asymmetric Graph Engine}\label{sec:sage}

Our main approach in \oursystem{} is to develop \SAMabbrev{}
techniques that perform \emph{no writes to the \largemem{}}. Using
these primitives lets us derive efficient parallel algorithms (i) whose
cost is independent of $\omega$, the asymmetry of the underlying
NVRAM technology, (ii) that do not contribute to NVRAM wearout or wear-leveling overheads, and (iii) that do not require on-the-fly recompression for compressed graphs. The surprising result of our experimental study
is that this strict discipline---to entirely avoid writes to the
\largemem{}---achieves state-of-the-art results in practice. This
discipline also enables storage optimizations (discussed in
Section~\ref{sec:exps}), and perhaps most importantly lends itself to
designing provably-efficient parallel algorithms that interact with
the graph through high-level primitives.

\myparagraph{Semi-Asymmetric \emap{}}
Our first contribution in \oursystem{} is a version of \emap{}
(Section~\ref{sec:prelims}) that
achieves improved efficiency in the \SAMabbrev{}. The issue with the
implementation of \emap{} used in Ligra, and subsequent systems (Ligra+
and Julienne) based on Ligra is that although it is work-efficient, it may use
significantly more than $O(n)$ space, violating the \SAMabbrev{} model.
In this paper, we design an improved implementation of \emap{} which
achieves superior performance in the \SAMabbrev{} model (described in
Section~\ref{subsec:blocking}). Our result is summarized by the
following theorem:

\begin{theorem}\label{thm:emap}
There is a \SAMabbrev{} algorithm for \emap{} given a \vset{} $U$
that runs in $O(\sum_{u \in U} \degree{u})$ work, $O(\log n)$ depth,
and uses at most $O(n)$ words of memory in the worst case.
\end{theorem}

\myparagraph{Semi-Asymmetric Graph Filtering}
An important primitive used by many parallel graph algorithms
performs \emph{batch-deletions} of edges incident to vertices over the
course of the algorithm. A batch-deletion operation is just a bulk
remove operation that logically deletes these edges from the graph.
These deletions are done to reduce the number of edges that must be
examined in the future. For example, four of the algorithms studied in
this paper---biconnectivity, approximate set cover, triangle counting,
and maximal matching---utilize this primitive.

In prior work in the shared-memory setting, deleted edges are
handled by actually removing them from the adjacency lists in the
graph.
In these algorithms, deleting edges is important for two reasons.
First, it reduces the amount of work done when edges incident to the
vertex are examined again, and second, removing the edges is important
to bound the theoretical efficiency of the resulting
implementations~\cite{dhulipala2017julienne, dhulipala18scalable}.
In the \SAMabbrev{}, however, deleting edges is expensive because it requires
writes to the \largemem{}.

In our \oursystem{} algorithms, instead of directly modifying the
underlying graph, we build an auxiliary data structure, which we refer
to as a \defn{\gfilter{}}, that efficiently supports updating a graph
with a sequence of deletions. The \gfilter{} data structure
can be viewed as a bit-packed representation of the original graph
that supports mutation. Importantly, this data structure fits into the
\smallmem{} of the
relaxed version of the \SAMabbrev{}. We formally define our data
structure and state our theoretical results in
Section~\ref{sec:filter}.

\input{inputs/intro_table}
\myparagraph{Efficient Semi-Asymmetric Graph Algorithms}
We use our new semi-asymmetric techniques to design efficient
semi-asymmetric graph algorithms for 18 fundamental graph problems.
In all but a few cases, the bounds are obtained by applying our new
semi-asymmetric techniques in conjunction with existing efficient
DRAM-only graph algorithms from Dhulipala et
al.~\cite{dhulipala18scalable}.  We summarize the \SAMabbrev{} work
and depth of the new algorithms designed in this paper in
Table~\ref{table:introtable}, and present the detailed results in
Section~\ref{sec:algorithms}.

\input{inputs/blocking}

\input{inputs/algorithm-casestudy}
\input{inputs/filtering}

\input{inputs/algorithm-bounds}

%% file: inputs/intro_table.tex
\newcommand{\STAB}[1]{\begin{tabular}{@{}c@{}}#1\end{tabular}}

\begin{table}[!ht]
\begin{center}
    \caption{Work and depth bounds of \oursystem{} algorithms in the
    \SAMabbrev{}. The GBBS Work column shows the work of GBBS
    algorithms converted to use NVRAM without taking advantage of the
    \smallmem{},
    and corresponds to the GBBS-NVRAM using libvmmalloc
    experiment (pink dashed-bars in Figure 7).  The theoretical
    performance for the GBBS-MemMode experiment (green dashed-bars in
    Figure 1) lies in-between the GBBS Work and Sage Work.
    The vertical text in the first column indicates the technique used
    to obtain the result in the \SAMabbrev{}: \emapchunk{} is the
    semi-asymmetric traversal in
    Section~\ref{subsec:blocking} and Filter is the Graph Filtering
    method in Section~\ref{sec:filter}.
    $^{\dag}$ denotes that our algorithm uses $O(n + m/\log
    n)$ words of memory.
    $^{\mathparagraph}$ denotes that our
    algorithm uses $O(n + m/\log n)$ words of memory in practice, but
    requires only $O(n)$ words of memory theoretically.
    $^{*}$ denotes that a bound holds
    in expectation and $^{\ddagger}$ denotes that a bound holds with
    high probability or \whp{} ($O(kf(n))$ cost with probability at
    least $1 - 1/n^{k}$).  $d_G$
    is the diameter of the graph, $\Delta$ is the maximum degree,
    $L=\min{(\sqrt{m},\Delta)}+\log^2\Delta\log n/\log\log n$, and
    $P_{\smb{it}}$ is the number of iterations of PageRank until
    convergence. We assume that $m=\Omega(n)$.
    }\label{table:introtable}
\vspace{0.03in}
\scalebox{0.85}{
\begin{tabular}{@{}c|l|l|l|l@{}}
  & \multicolumn{1}{l|}{\textbf{Problem}} & \textbf{GBBS Work} & \textbf{\oursystem{} Work} & \textbf{Sage Depth} \\ \hline
    \multirow{11}{*}{\STAB{\rotatebox[origin=c]{90}{\emapchunk{}\ \ \ }}}
    & {\bf Breadth-First Search                          }  & $O(\omega m)$                    & $O(m)$                     & $O(\mathsf{d}_G \log n)$ \\
    & {\bf Weighted BFS                 }  & $O(\omega m)^{*}$                & $O(m)^{*}$                 & $O(\mathsf{d}_G \log n)$$^\ddagger$ \\
    & {\bf Bellman-Ford                 }  & $O(\omega \mathsf{d}_G m)$  & $O(\mathsf{d}_G m)$   &      $O(\mathsf{d}_G \log n)$ \\
    & {\bf Single-Source Widest Path            }  & $O(\omega \mathsf{d}_G m)$  & $O(\mathsf{d}_G m)$        & $O(\mathsf{d}_G \log n)$ \\
    & {\bf Single-Source Betweenness            }  & $O(\omega m)$                    & $O(m)$                     & $O(\mathsf{d}_G \log n)$ \\
    & {\bf $O(k)$-Spanner               }  & $O(\omega m)^{*}$                & $O(m)^{*}$                 & $O(k\log n)$$^\ddagger$ \\
    & {\bf LDD                          }  & $O(\omega m)^{*}$                & $O(m)^{*}$                 & $O(\log^2 n)$$^\ddagger$\\
    & {\bf Connectivity        }  & $O(\omega m)^{*}$                & $O(m)^{*}$                 & $O(\log^3 n)$$^\ddagger$ \\
    & {\bf Spanning Forest     }  & $O(\omega m)^{*}$                & $O(m)^{*}$                 & $O(\log^3 n)$$^\ddagger$ \\
    & {\bf Graph Coloring               }  & $O(\omega m)^{*}$                & $O(m)^{*}$                 & $O(\log n + L \log \Delta)^*$ \\
    & {\bf Maxmial Independent Set                          }  & $O(\omega m)^{*}$                & $O(m)^{*}$                 & $O(\log^2 n)$$^{\ddagger}$ \\
    \midrule
        \multirow{2}{*}{\STAB{\rotatebox[origin=c]{90}{Both}}}
    & {\bf Biconnectivity$^{\mathparagraph}$     }   & $O(\omega m)^{*}$                & $O(m)^{*}$                 & $O(\mathsf{d}_G\log n+ \log^{3} n)^{\ddagger}$  \\
    & {\bf Approximate Set Cover$^{\dag}$      }  & $O(\omega m)^{*}$                & $O(m)^{*}$                 & $O(\log^{3} n)$$^{\ddagger}$ \\
    \midrule
    \multirow{2}{*}{\STAB{\rotatebox[origin=c]{90}{Filter}}}
    & {\bf  Triangle Counting$^{\dag}$  }  & $O(\omega (m + n) + m^{3/2})$               & $O(m^{3/2})$               & $O(\log n)$ \\
    & {\bf Maximal Matching$^{\dag}$   }   & $O(\omega m)^{*}$                & $O(m)^{*}$                  & $O(\log^3 m)$$^{\ddagger}$ \\
    \midrule
    & {\bf PageRank Iteration           }  & $O(m + \omega n)$                    & $O(m)$                     & $O(\log n)$ \\
    & {\bf PageRank                     }  & $O(P_{\emph{it}}(m + \omega n))$ & $O(P_{\emph{it}}m)$  & $O(P_{\emph{it}}\log n)$ \\
    & {\bf $k$-core                     }  & $O(\omega m)^{*}$                & $O(m)^{*}$                 & $O(\rho \log n)$$^{\ddagger}$ \\
    & {\bf Approximate Densest Subgraph           }  & $O(\omega m)$                    & $O(m)$                     & $O(\log^2 n)$ \\
  \end{tabular}
}
\end{center}
  \par
\end{table}

%% file: inputs/blocking.tex
\subsection{Semi-Asymmetric Graph Traversal}\label{subsec:blocking}
Our first technique is a cache-friendly and memory-efficient sparse
\emap{} primitive designed for the \SAMabbrev{}. This technique is
useful for obtaining \SAMabbrev{} algorithms for many of the problems
studied in this paper.  Graph traversals are a basic graph primitive,
used throughout many graph algorithms~\cite{CLRS, ShunB2013}. A
graph traversal starts with a frontier (subset) of seed vertices. It
then runs a number of iterations, where in each iteration, the edges
incident to the current frontier are explored, and vertices in this
neighborhood are added to the next frontier based on some user-defined
conditions.

\subsubsection{Existing Memory-Inefficient Graph Traversal}
Ligra implements the direction-optimization proposed by
Beamer~\cite{beamer12direction}, which runs either a \emph{sparse}
(push-based) or \emph{dense} (pull-based) traversal, based on the
number of edges incident to the current frontier. The sparse traversal
processes the out-edges of the current frontier to generate the
next frontier. The dense traversal processes the in-edges of all
vertices, and checks whether they have a neighbor in the current
frontier. Ligra uses a threshold to select a method, which by default
is a constant fraction of $m$ to ensure work-efficiency.

The dense method is memory-efficient---theoretically, it only requires
$O(n)$ bits to store whether each output vertex is on the next
frontier. However, the sparse method can be memory-inefficient because
it allocates an array with size proportional to the number of edges
incident to the current frontier, which can be up to $O(m)$.  In the
\SAMabbrev{}, an array of this size can only be allocated in the
\largemem{}, so the traversal is inefficient. This is also true for
the real graphs and machines that we tested in this paper.

The GBBS algorithms~\cite{dhulipala18scalable} use a \emph{blocked}
sparse traversal, referred to as \emapblock{}, that improves the
cache-efficiency of parallel graph traversals by only writing to as
many cache lines as the size of the newly generated frontier. This
technique is not memory-efficient, as it allocates an intermediate
array with size proportional to the number of edges incident to the
current frontier, which can be up to $O(m)$ in the worst-case.

\subsubsection{Memory-Efficient Traversal: \emapchunk{}}
In this paper, we present a chunk-based approach that improves the
memory-efficiency of the sparse (push-based) \emap{}. Our approach,
which we refer to as \emapchunk{},
achieves the same cache performance as the \emapblock{} implementation
used in GBBS~\cite{dhulipala18scalable}, but
significantly improves the intermediate memory usage of the approach.
We provide the details of our algorithm and its pseudocode in
Appendix~\ref{apx:blocking} and \ref{apx:emapchunk_mem}.

\myparagraph{Our Algorithm}
The high-level idea of our algorithm is as follows. It first
divides the edges that are to-be traversed into grouped units of work.
This is done based on the underlying group size, $g$, of the graph,
which is set to the average degree $\davg$.
The edges incident to each vertex are
partitioned into groups based on $g$. The algorithm then performs a
work-assignment phase, which statically load-balances the work over
the incident edges to $O(P)$ virtual threads. Next, in parallel for
each virtual thread, it processes the edges assigned to the thread.
For each group, it uses a thread-local allocator to obtain a
\emph{chunk} that is ensured to have sufficient memory to store the
output of mapping over the edges in the group. The chunks are stored
in thread-local vectors. Upon completion of processing all edges
incident to the \vset{}, the algorithm aggregates all chunks stored in
the thread-local vectors and uses a prefix-sum and a parallel copy to
store the output neighbors contiguously in a single flat array.
The overall work of the procedure is
$O(\sum_{u\in U}\degree{u})$ where $U$ is the input \vset{}, and
the depth is $O(\log n)$, which match the work and depth bounds of the
previous \emapblock{} implementation. We note that for compressed
graphs, we must set the underlying group size of the graph to the average
degree, which makes the depth $O(\log n + \davg{})$ (please see
Appendix~\ref{apx:blocking} for details).  Our
algorithm also obtains the same cache-efficiency as \emapblock{},
while improving the memory usage of the algorithm to $O(n)$ words.

%% file: inputs/algorithm-casestudy.tex
\input{inputs/bfs_figure}
\subsubsection{Case Study: Breadth-First Search}\label{sec:casestudy}

\myparagraph{Algorithm}
Figure~\ref{fig:bfs} provides the full \oursystem{} code used for our
implementation of BFS. The algorithm outputs a BFS-tree, but can
trivially be modified to output shortest-path distances from the
source to all reachable vertices.
The user first imports the \oursystem{} library
(Line~\ref{line:importsage}).  The definition of \textsc{BFSFunc} defines
the user-defined function supplied to \emap{}
(Lines~\ref{line:bfsstart}--\ref{line:bfsend}).  The main algorithm,
\textsc{BFS}, is templatized over a graph type
(Line~\ref{line:template}).
\hide{
One of the advantages of our
implementation is that the \emph{same} BFS implementation works
regardless of whether the underlying graph is an ordinary immutable
graph, or a graph filter.}
The BFS code first initializes the parent
array \textsc{P} (Line~\ref{line:initparents}), sets the parent of the source
vertex to itself (Line~\ref{line:setsrcparent}), and initializes the
first frontier to contain just the parent
(Line~\ref{line:setfrontier}). It then loops while the frontier is
non-empty (Lines~\ref{line:whilestart}--\ref{line:whileend}), and
calls \emapchunk{} in each iteration of the while loop (Line~\ref{line:emapchunk}).

The function supplied to \emapchunk{} is \textsc{BFSFunc}
(Lines~\ref{line:bfsstart}--\ref{line:bfsend}), which contains two
implementations of update based on whether a sparse or dense traversal
is applied (\textsc{update} and \textsc{updateAtomic} respectively),
and the function \textsc{cond} indicating whether a neighbor should be
visited. This logic is identical to the update function used in BFS in
Ligra, and we refer the interested reader to Shun and
Blelloch~\cite{ShunB2013} for a detailed explanation.

\myparagraph{\SAMabbrev{}: Work-Depth Analysis}
The work is calculated as follows. First, the work of initializing the
parent array, and constructing the initial frontier is just $O(n)$.
The remaining work is to apply \emap{} across all rounds. To bound
this quantity, first observe that each vertex, $v$, processes its
out-edges at most once, in the round where it is contained in
$\codevar{Frontier}$ (if other vertices try to visit $v$ in subsequent
rounds notice that the \textsc{cond} function will return
\emph{false}). Let $R$ be the set of all rounds run by the algorithm,
$W_{\emapchunk{}}(r)$ be the work of \emapchunk{} on the $r$-th round,
and $U_{r}$ be the set of vertices in $\codevar{Frontier}$ in the
$r$-th round. Then, the work is
$\sum_{r \in R} W_{\emapchunk{}}(r) = \sum_{r \in R} \sum_{u \in U_r}
\degree{u} = O(m)$.

The depth to initialize the parents array is $O(\log n)$, and the
depth of each of the $r$ applications of \emapchunk{} is $O(\log n)$
by Theorem~\ref{thm:emap}. Thus, the overall depth is $O(r
\log n) = O(\mathsf{d}_{G} \log n)$. The \smallmem{} space
used for the parent array is $O(n)$ words, and the maximum space used over
all \emapchunk{} calls is $O(n)$ words by Theorem~\ref{thm:emap}.
This proves the following theorem:
\begin{theorem}\label{thm:semiasymbfs}
There is a \SAMabbrev{} algorithm for breadth-first search that runs in
$O(m)$ work, $O(\mathsf{d}_{G} \log n)$ depth, and uses only $O(n)$
words of \smallmem{}.
\end{theorem}

%% file: inputs/bfs_figure.tex
\definecolor{light-gray}{gray}{0.95}
\definecolor{dark-gray}{gray}{0.25}

\begin{figure*}[!ht]
\centering
\begin{minted}[fontsize=\small,linenos=true,numbers=left,autogobble,xleftmargin=0.2\textwidth,xrightmargin=0.2\textwidth,frame=single,escapeinside=||]{cpp}
#include "sage.h"|\label{line:importsage}|
#include <limits>
template <class W, class Int>|\label{line:bfsstart}|
struct BFSFunc {
  sequence<Int>& P;
  Int max_int;
  BFSFunc(sequence<Int>& P) : P(P) {
    max_int = std::numeric_limits<Int>::max();}
  bool update(Int s, Int d, W w) {
    if (P[d] == max_int) {
      P[d] = s;
      return 1;
    }
    return 0;
  }
  bool updateAtomic(Int s, Int d, W w) {
    return (CAS(&P[d], max_int, s));
  }
  bool cond(Int d) { return (P[d] == max_int); }
}; |\label{line:bfsend}|
template <class Graph, class Int> |\label{line:template}|
sequence<Int> BFS(Graph& G, Int src) {
  using W = typename Graph::weight_type;
  Int max_int = std::numeric_limits<Int>::max();
  auto P = sequence<Int>(G.n, max_int);|\label{line:initparents}|
  P[src] = src;|\label{line:setsrcparent}|
  auto frontier = vertexSubset(G.n, src); |\label{line:setfrontier}|
  while (!frontier.isEmpty()) { |\label{line:whilestart}|
    auto F = BFSFunc<W, Int>(P);
    frontier = edgeMapChunked(G, frontier, F); |\label{line:emapchunk}|
  }|\label{line:whileend}|
  return P;
}
\end{minted}
\vspace{-0.05in}
\caption{\revised{Code for Breadth-First Search in \oursystem{}.}}
\label{fig:bfs}
\end{figure*}

%% file: inputs/filtering.tex
\subsection{Semi-Asymmetric Graph Filtering}\label{sec:filter}

\oursystem{} provides a high-level filtering interface that captures
both the current implementation of filtering in GBBS, as well as the
new mutation-avoiding implementation described in this paper.
The
interface provides functions for creating a new \gfilter{}, filtering
edges from a graph based on a user-defined predicate, and a function
similar to \emap{} which filters edges incident to a subset of
vertices based on a user-defined predicate.
Since edges incident to a vertex can be deleted over the course of the
algorithm by using a \gfilter{}, we call edges that are currently part
of the graph represented by the \gfilter{} as \defn{active} edges.

We first discuss a semantic issue that arises when filtering graphs.
Suppose the user builds a filter $G_f$ over a symmetric graph $G$. If
the filtering predicate takes into account the directionality of the
edge, then the resulting graph filter can become directed, which is
unlikely to be what the user intends. Therefore, we designed the
constructor to have the user explicitly specify this decision by
indicating whether the user-defined predicate is symmetric or
asymmetric, which results in either a symmetric or asymmetric
graph filter.

The filtering interface is defined as follows:

\begin{itemize}[topsep=0pt,itemsep=0ex,partopsep=0ex,parsep=1ex, leftmargin=*]
  \item{\textbf{\makefilter{}}($G : \mathsf{Graph}$,} \\
    \hspace*{2.0cm}$P : \mathsf{edge} \mapsto \mathsf{bool}$, $S$ : $\mathsf{bool}$) :
    $\mathsf{\gfilter{}}$\\
Creates a \gfilter{} $G_{f}$ for the immutable graph $G$
with respect to the user-defined predicate $P$, and $S$, which
indicates whether the filter is symmetric or asymmetric.

\item{\textbf{\filteredges{}}($G_f : \mathsf{\gfilter{}}$}) : $\mathsf{int}$ \\
Filters all active edges in $G_f$ that do not satisfy the predicate
$P$ from $G_f$. The function mutates the supplied \gfilter{}, and
returns the number of edges remaining in the \gfilter{}.

\item \textbf{\emappack{}}($G_f : \mathsf{\gfilter{}}$, \\
  \hspace*{2.3cm} $S : \mathsf{\vset{}}) :  \mathsf{\vsetaug{}}$ \\
Filters edges incident to $v \in S$ that do not satisfy the
predicate $P$ from $G_{f}$. Returns a \vset{} on the same vertex set as $S$, where each
vertex is augmented with its new degree in $G_f$.
\end{itemize}

\subsubsection{Graph Filter Data Structure}\label{sec:filterds}
For simplicity, we describe the symmetric version of the graph filter
data structure. The asymmetric filter follows naturally by using two
copies of the data structure described below, one for the in-edges and
one for the out-edges.

We first review how edges are represented in \oursystem{}. In the
(uncompressed) CSR format, the neighbors of a vertex are stored
contiguously in an array. If the graph is compressed using one of the
parallel compression methods from Ligra+~\cite{shun2015ligraplus}, the
incident edges are divided into a number of \emph{compression blocks},
where each block is sequentially encoded using a difference-encoding
scheme with variable-length codes. Each block must be sequentially
decoded to retrieve the neighbor IDs within the block, but by choosing
an appropriate block size, the edges incident to a high-degree vertex
can be traversed in parallel across the blocks.

\input{inputs/figure-filter}

The graph filter's design \emph{mirrors the CSR representation}
described above. The design of our structure is inspired by similar
bit-packed structures, most notably the cuckoo-filter by Eppstein et
al.~\cite{Eppstein2017}.  Figure~\ref{fig:filter} illustrates the
graph filter for the following description.

\myparagraph{Definition}
The graph data is stored in the compressed sparse row (CSR) format on
NVRAM, and is read-only. Each vertex's incident edges are logically
divided into blocks of size \filterbs{}, the \defn{filter block size},
which is the provided block size rounded up to the next multiple of
the number of bits in a machine word, inclusive (64 bits on modern
architectures and $\log n$ bits in theory). In
Figure~\ref{fig:filter}, \filterbs{} = $2$. For compressed
graphs, this block size is always equal to the compression block size
(and thus, both must be tuned together).

The filter consists of blocks corresponding to a subset of the logical
blocks in the edges array, and is stored in DRAM.  Each vertex stores
a pointer to the start of its blocks, which are stored contiguously.
For each block, the filter stores $\filterbs{}$ many bits, where the
bits correspond one-to-one to the edges in the block. Each block also
stores two words of metadata: (i) the \emph{original block-ID}
in the adjacency list that the block corresponds to, and (ii) the
\emph{offset}, which stores the number of active edges before this
block. The original block-IDs are necessary because over the course of
the algorithm, only a subset of the original blocks used for a vertex
may be currently present in the graph filter, and the data structure
must remember the original position of each block. The offset is
needed for graph primitives which copy all active edges incident to a
vertex into an array with size proportional to the degree of the
vertex.

The overall graph filter structure thus consists of blocks of bitsets
per vertex. It stores the per-vertex blocks contiguously, and stores
an offset to the start of each vertex's blocks. It also stores each
vertex's current degree, as well as the number of blocks in the vertex
structure. Finally, the structure stores an additional $n$ bits of
memory which are used to mark vertices as dirty.

\subsubsection{Algorithms}
\hide{
We now describe how the graph filter primitives are implemented in
\oursystem{}, and analyze their costs. Our algorithms support both
uncompressed graphs, and graphs compressed using the parallel encoding
scheme from Ligra+~\cite{shun2015ligraplus}.
}

\myparagraph{\textsc{MakeFilter}} To create a graph filter, the algorithm first
computes the number of blocks that each vertex requires, based on
\filterbs{}, and writes the space required per vertex into an array.
Next, it prefix sums the array, and allocates the required $O(m)$ bits of
memory contiguously. It then initializes the per-vertex blocks in
parallel, setting all edges as active (their corresponding
bit is set to 1). Finally, it allocates an array of $n$ per-vertex
structures storing the degree, offset into the bitset structure
corresponding to the start of the vertex's blocks, and the number of
blocks for that vertex. Lastly, it initializes per-vertex dirty bits
to $\mathsf{false}$ (not dirty) in parallel.

The overall work to create the filter is $O(m)$ and the overall depth
is $O(\log n + \filterbs{})$, because a block is processed
sequentially. If the user specifies that the initially supplied
predicate returns $\mathsf{false}$ for some edges, the implementation
calls \filteredges{} (described below), which runs within the same
work and depth bounds.

\myparagraph{\textsc{PackVertex}} Next, we describe an algorithm to
pack out the edges incident to a vertex given a predicate $P$.
This algorithm is an internal primitive in \oursystem{} and is not
exposed to the user. The algorithm first maps over all blocks
incident to the vertex in parallel.

For each block, it finds all active bits in the block, reads the edge
corresponding to the active bit and applies the predicate $P$,
unsetting the bit iff the predicate returns $\mathsf{false}$. If the
bit for an edge $(u, v)$ is unset, the algorithm marks the dirty bit
for $v$ to true if needed.
The algorithm maintains a count of how many bits are still active
while processing the block, and stores the per-block counts in an
array of size equal to the number of blocks. Next, it
performs a reduction over this array to compute the number of blocks
with at least one active edge. If this value is less than a constant
fraction of the current number of blocks incident to the vertex, the
algorithm filters out all of these blocks with no active elements, and
packs the remaining blocks contiguously in the same memory, using a
parallel filter over the blocks.
The algorithm then updates the offsets for all blocks using a prefix sum. Finally, the algorithm updates the vertex
degree and number of currently active blocks incident to the vertex.

The overall work is $O(A \cdot (\filterbs{}/\log n) +
d_{\emph{active}}(v))$ and the depth is $O(\log n + \filterbs{})$,
where $A$ is the number of non-empty blocks corresponding to $v$ and
$d_{\emph{active}}(v)$ is the number of active edges incident to
vertex $v$.

\myparagraph{\emappack{}} The \emappack{} primitive is implemented by
applying \textsc{PackVertex} to each vertex in the \vset{} in
parallel. It then updates the number of active edges by performing a
reduction over the new vertex degrees in the \vset{}. The overall work
is the sum of the work for packing out each vertex in the \vset{},
$S$, which is $O(A\cdot (\filterbs{}/\log n) + \sum_{v\in
S}(1+d_{\emph{active}}(v)))$, and the depth is $O(\log n +
\filterbs{})$, where $A$ is the number of non-empty blocks
corresponding to all $v \in S$.

\myparagraph{\filteredges{}} The \filteredges{} primitive uses the \emappack{}, providing a \vset{} containing all
vertices. The work is $O(n + A\cdot (\filterbs{}/\log n) +
|E_{active}|)$, and the depth is $O(\log n + \filterbs{})$, where $A$
is the number of non-empty blocks in the graph and $E_{active}$ is the
set of active edges represented by the graph filter.

\hide{
\myparagraph{Vertex Primitives} When calling vertex primitives on a graph
filter, such as accessing the degree of the vertex, or mapping over
its incident active edges, we first check whether the vertex is marked
dirty. If so, we pack it out using the algorithm described above
before performing the operation.
}

\subsubsection{Implementation}

\myparagraph{Optimizations}
We use the widely available \textsc{tzcnt} and \textsc{blsr} x86
intrinsics to accelerate block processing. Each block is logically
divided into a number of machine words, so we consider processing a
single machine word. If the word is non-zero, we create a temporary
copy of the word, and loop while this copy is non-zero. In each
iteration, we use \textsc{tzcnt} to find the index of the next lowest
bit, and clear the lowest bit using \textsc{blsr}. Doing so allows us
to process a block with $q$ words and $k$ non-zero bits in $O(q + k)$
instructions.

We also implemented intersection primitives, which are used in our
triangle counting algorithm based on the decoding implementation
described above. For compressed graphs, since we may have to decode an
entire compressed block to fetch a single active edge, we immediately
decompress the entire block and store it locally in the iterator's
memory. We then process the graph filter's bits word-by-word using the
intrinsic-based algorithm described above.

\myparagraph{Memory Usage}
The overall memory requirement of a \gfilter{} is $3n$ words to store the degrees,
offsets, and number of blocks, plus $O(m)$ bits to store the bitset data and
the metadata. The metadata increases the memory usage by a constant
factor, since \filterbs{} is at least the size of a machine word, and
so the metadata stored per block can be amortized against the bits
stored in the block. The overall memory usage is therefore $O(n +
m/\log n)$ words of memory.
For our uncompressed inputs, the size of the graph filter is
4.6--8.1x smaller than the size of the uncompressed graph. For our
compressed inputs, the size of the filter is 2.7--2.9x smaller
than the size of the compressed graph.

%% file: inputs/figure-filter.tex
\begin{figure}
\begin{center}
\includegraphics[width=0.65\columnwidth]{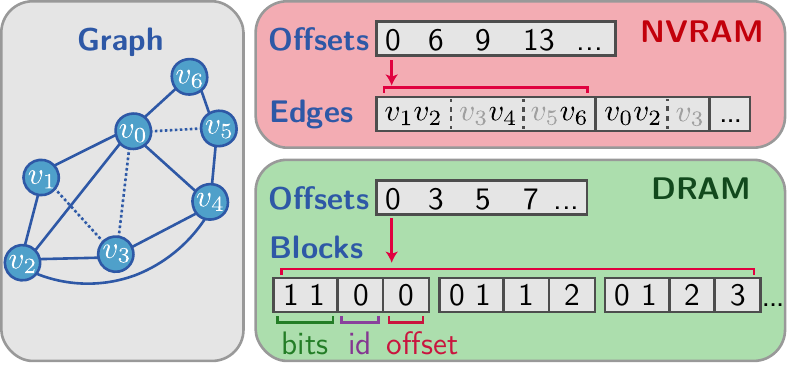}
\caption{\label{fig:filter}
%This figure illustrates our graph filter data structure, and is
%described in detail in Section~\ref{sec:filterds}.
The graph data is stored in the compressed sparse row (CSR) format on
NVRAM, and is read-only. Each vertex's incident edges are logically
divided into blocks of size \filterbs{} each (\filterbs{} = 2 here).
The filter is structured similarly to the CSR format and is stored in
DRAM. It consists of blocks corresponding to a subset of the logical
blocks in the edges array. Each vertex stores a pointer
to the start of its blocks, which are stored contiguously.  Each
filter block stores bits representing the status of the edges in the logical block
it corresponds to. In the figure, deleted edges are marked with dotted
lines, and correspond to bits set to 0 in the filter. Each block also
stores an offset storing the number of active edges (edges
with bits set to 1) in all preceding blocks for this vertex, as well
as its original block-id, which are both used by our algorithms.  When
an edge is deleted, its corresponding bit is set to 0, and the offsets
of the blocks for the vertex are updated accordingly.
The original graph data in stored in the CSR format and is stored on NVRAM and is read-only.
On DRAM, we maintain the filter structure that consists of blocks.
Each block corresponds to \filterbs{} many edges in a consecutive
range, and stores \filterbs{} many bits for these edges. In addition
it stores an offset and the original block-id for each block, which
are used by our algorithms.
When an edge is deleted, its corresponding bit is set to 0, and the
offsets of the blocks for the vertex are updated accordingly.
Once a constant fraction of blocks are empty (e.g., the gray block), we physically delete all the empty blocks from the filter structure to guarantee work-efficiency.
All writes are on DRAM.
}
\vspace{-1em}
\end{center}
\end{figure}

%% file: inputs/algorithm-bounds.tex
\subsection{Semi-Asymmetric Graph Algorithms}\label{sec:algorithms}

We now describe \oursystem{}'s efficient parallel graph algorithms in
the \SAMabbrev{} model. Our results and theoretical
bounds are summarized in Table~\ref{table:introtable}. The bounds are
obtained by combining efficient in-memory algorithms in our prior
work~\cite{dhulipala18scalable} with the new semi-asymmetric
techniques designed in Sections~\ref{subsec:blocking} and
\ref{sec:filter}.
%Due to space constraints,
We elide some of the details of the theoretical results by describing
how the results are obtained based on the proofs
from Dhulipala et al.~\cite{dhulipala18scalable}.
Specifications of the graph problems and additional algorithm details can be
found in Appendix~\ref{apx:algorithmspecs}.

\subsubsection{Shortest Path Problems}\label{sec:shortestpath}

\myparagraph{Algorithms}
We consider six shortest-path problems in this paper:
\defn{breadth-first search (BFS)}, \defn{integral-weight SSSP (wBFS)},
\defn{general-weight SSSP (Bellman-Ford)}, \defn{single-source
betweenness centrality}, \defn{single-source widest path}, and
\defn{$\bm{O(k)}$-spanner}. Our BFS, Bellman-Ford,
and betweenness centrality implementations are based on those in
Ligra~\cite{ShunB2013}, and our wBFS implementation is based on the one in
Julienne~\cite{dhulipala2017julienne}. We provide two
implementations of the single-source widest path algorithm, one based
on Bellman-Ford, and another based on the wBFS implementation from
Julienne~\cite{dhulipala2017julienne}. An $O(k)$-spanner is a subgraph
that preserves shortest-path distances within a factor of $O(k)$.  Our
$O(k)$-spanner implementation is based on an
algorithm by Miller et al.~\cite{miller2015spanners}.

\myparagraph{Efficiency in the \SAMabbrev{}}
Our theoretical bounds for these problems in the \SAMabbrev{} are
obtained by using the \emapchunk{} primitive
(Section~\ref{subsec:blocking}) for performing sparse graph
traversals, because all of these algorithms can be expressed as
iteratively performing \emapchunk{} over subsets of vertices. The
proofs are similar to the proof we provide for BFS in
Section~\ref{sec:casestudy} and rely on Theorem~\ref{thm:emap}. Note
that the bucketing data structure used in
Julienne~\cite{dhulipala2017julienne} requires only $O(n)$ words of
space to bucket vertices, and thus automatically fits in the
\SAMabbrev{} model.
The Miller et al. construction builds an $O(k)$-spanner with
size $O(n^{1+1/k})$, and runs in $O(m)$ expected work and $O(k\log n)$
depth \whp{}. Our implementation in \oursystem{} runs our
low-diameter decomposition algorithm, which is efficient in the
\SAMabbrev{} as we describe below. We set $k$ to be $\Theta(\log n)$,
which results in a spanner with size $O(n)$.

\subsubsection{Connectivity Problems}\label{sec:connectivity}

\myparagraph{Algorithms}
We consider four connectivity problems in this paper:
\defn{low-diameter decomposition (LDD)}, \defn{connectivity},
\defn{spanning forest}, and \defn{biconnectivity}. Our implementations
are extensions of the implementations provided in
GBBS~\cite{dhulipala18scalable}.
%, and due to space constraints we refer
%the reader to that paper for detailed descriptions of the problems and
%of our provably-efficient in-memory algorithms.

%\myparagraph{Efficiency in the \SAMabbrev{}}
%First, we replace the calls to \emapblock{} in each algorithm with
%calls to \emapchunk{}, which ensures that the graph traversal step
%uses $O(n)$ words of \smallmem{} using Theorem~\ref{thm:emap}. This
%modification results in \SAMabbrev{} algorithms for LDD.
%For the other connectivity-like algorithms that use LDD, namely
%connectivity, spanning forest, and biconnectivity, we run the
%algorithms in the relaxed \SAMabbrev{} model. We use the extra space
%provided in the relaxed model to apply LDD with $\beta = O(1/\log^{c}{n})$
%for an integer $c > 0$, so that we can explicitly build the
%inter-cluster graph induced by the LDD, which contains $O(m/\log n)$
%edges in expectation. We can obtain the same space bounds in the
%relaxed model in the worst case for our connectivity, spanning forest,
%and biconnectivity algorithms without affecting the work and depth of
%the algorithms (see Appendix~\ref{sec:connectivityproblems} for details). Lastly, we
%note that our biconnectivity implementation uses the graph filtering
%structure to optimize a call to connectivity which runs on the input
%graph, with a large subset of the edges removed.

\myparagraph{Efficiency in the \SAMabbrev{}}
First, we replace the calls to \emapblock{} in each algorithm with
calls to \emapchunk{}, which ensures that the graph traversal step
uses $O(n)$ words of \smallmem{} using Theorem~\ref{thm:emap}. This
modification results in \SAMabbrev{} algorithms for LDD.
For the other connectivity-like algorithms that use LDD, namely
connectivity, spanning forest, biconnectivity, we use the improved
analysis of LDD provided in~\cite{miller2015spanners} to argue that
the number of inter-cluster edges after applying LDD with $\beta =
O(1)$ is $O(n)$ in expectation. Thus, the graph on the inter-cluster
edges can be built in \smallmem{}. We provide the full details
in Appendix~\ref{sec:connectivityproblems}.
We also obtain the same space bounds in the worst case for our
connectivity, spanning forest, biconnectivity, and $O(k)$-spanner
algorithms without affecting the work and depth of the algorithms
(see Appendix~\ref{sec:connectivityproblems} for details).
Lastly, we note that although
this results in only $O(n)$ words of \smallmem{} theoretically, in practice
our implementation of biconnectivity instead uses the graph filtering
structure to optimize a call to connectivity that runs on the input
graph, with a large subset of the edges removed.

\subsubsection{Covering Problems}\label{sec:covering}
\myparagraph{Algorithms}
We consider four covering problems in this paper: \defn{maximal
independent set (MIS)}, \defn{maximal matching}, \defn{graph
coloring}, and \defn{approximate set cover}. All of our
implementations are extensions of our previous work in
GBBS~\cite{dhulipala18scalable}.

\myparagraph{Efficiency in the \SAMabbrev{}}
For MIS and graph coloring, we derive \SAMabbrev{} algorithms
by applying our \emapchunk{} optimization because other than graph
traversals, both algorithms already use $O(n)$ words of \smallmem{}.
Both maximal matching and approximate set cover use our graph
filtering technique to achieve immutability and reduced memory usage
without affecting the theoretical bounds of the algorithms. We provide
more details about our maximal matching algorithm in
Appendix~\ref{sec:coveringproblems}. For set cover, our new
bounds for filtering match the bounds on filtering used in the GBBS
code which mutates the underlying graph, and so our implementation
also computes a $(1+\epsilon)$-approximate set cover in $O(m)$
expected work and $O(\log^{3} n)$ depth \whp{}.

\subsubsection{Substructure Problems}\label{sec:substructure}
\myparagraph{Algorithms}
We consider three substructure-based problems in this paper:
\defn{$\bm{k}$-core},
\defn{approximate densest subgraph} and \defn{triangle counting}. Substructure
problems are fundamental building blocks for community detection and
network analysis (e.g.,~\cite{bonchi2019distance, huang2014querying, 
li2013efficient, sariyuce2018local, sariyuce2015finding, 
wang2012truss}). Our $k$-core and triangle-counting
implementations are based on the implementation from
GBBS~\cite{dhulipala18scalable}.

\myparagraph{Efficiency in the \SAMabbrev{}}
For the $k$-core algorithm to use $O(n)$ words of \smallmem{}, it
should use the fetch-and-add based implementation of $k$-core, which
performs atomic accumulation in an array in order to update the
degrees. However, the fetch-and-add based implementation performs
poorly in practice, where it incurs high contention to update the
degrees of vertices incident to many removed
vertices~\cite{dhulipala18scalable}. Therefore, in practice we use a
\emph{histogram-based} implementation, which always runs faster than
the fetch-and-add based implementation (the histogram primitive is
fully described in~\cite{dhulipala18scalable}). In this paper, we
implemented a \emph{dense} version of the histogram routine, which
performs reads for all vertices in the case where the number of
neighbors of the current frontier is higher than a threshold $t$. The
work of the dense version is $O(m)$. Using $t=m/c$ for
some constant $c$ ensures work-efficiency, and results in low memory
usage for sparse calls in practice.
Our approximate densest subgraph algorithm is similar to our $k$-core
algorithm, and uses a histogram to accelerate processing the removal
of vertices. The code uses the dense histogram optimization
described above.
Our triangle counting implementation uses the graph
filter structure to orient edges in the graph from lower degree to
higher degree.

\subsubsection{Eigenvector Problems}\label{sec:eigenvector}
\myparagraph{Algorithms}
We consider the \defn{PageRank} algorithm, designed to rank the
importance of vertices in a graph~\cite{brin1998pagerank}.
Our PageRank
implementation is based on the implementation from Ligra.

\myparagraph{Efficiency in the \SAMabbrev{}}
We optimized the Ligra implementation to improve the depth of the
algorithm. The implementation from Ligra runs dense iterations, where
the aggregation step for each vertex (reading its neighbor's PageRank
contributions) is done sequentially. In \oursystem{}, we implemented a
reduction-based method that reduces over these neighbors using a
parallel reduce.  Therefore, each iteration of our implementation
requires $O(m)$ work and $O(\log n)$ depth. The overall work is
 $O(P_{it}\cdot m)$
and depth is $O(P_{it} \log n)$, where $P_{it}$ is the
number of iterations required to run PageRank to convergence with a
convergence threshold of $\epsilon=10^{-6}$.

%% file: inputs/experiments.tex
\section{Experiments}\label{sec:exps}

\revised{
\myparagraph{Overview of Results}
After describing the experimental setup (Section~\ref{sec:setup}),
we show the following main experimental results:

\begin{itemize}[topsep=0pt,itemsep=0pt,parsep=0pt,leftmargin=8pt]
  \item {\bf Section~\ref{sec:graphstorage}:}
  Our NUMA-optimized graph storage approach outperforms naive (and
  natural) approaches by 6.2x.

  \item {\bf Section~\ref{subsec:scalability}:} \oursystem{}
  achieves between 31--51x speedup for shortest path
  problems, 28--53x speedup for connectivity problems, 16--49x speedup for covering
  problems, 9--63x speedup for substructure problems and 42--56x
  speedup for eigenvector problems.

  \item {\bf Section~\ref{subsec:usvsinmem}:} Compared to existing
  state-of-the-art DRAM-only graph analytics, \oursystem{} run on
  NVRAM is 1.17x faster on average on our largest graph that
  fits in DRAM. \oursystem{} run on NVRAM is only 5\% slower on
  average than when run entirely in DRAM.

  \item {\bf Section~\ref{subsec:usvsothernvm}:} We study how
  \oursystem{} compares to other NVRAM approaches for graphs that are
  larger than DRAM. We find that \oursystem{} on \nvm{} using
  \appdirectmode{} is 1.94x faster on average than the
  recent state-of-the-art Galois codes~\cite{Gill2019} run using
  \memorymode{}. Compared to GBBS codes run using \memorymode{},
  \oursystem{} is 1.87x faster on average across all 18
  problems.

  \item {\bf Section~\ref{subsec:usvssemiexternal}:} We compare
  \oursystem{} with existing state-of-the-art semi-external memory
  graph processing systems, including FlashGraph, Mosaic, and
  GridGraph and find that our times are 9.3x, 12x, and 8024x faster on
  average, respectively.
\end{itemize}
}

\subsection{Experimental Setup}\label{sec:setup}
\subsubsection{Machine Configuration} We run our experiments on a
48-core, 2-socket machine (with two-way hyper-threading) with $2\times
2.2\mbox{Ghz}$ Intel 24-core Cascade Lake processors (with 33MB L3
cache) and 375GB of DRAM. The machine has 3.024TB of
\nvm{} spread across 12 252GB DIMMs (6 per socket). All of our \emph{speedup}
numbers report running times on a single thread (T1) divided by running times
on \emph{48-cores with hyper-threading} (T96). Our programs are compiled
with the \texttt{g++} compiler (version 7.3.0) with the \texttt{-O3}
flag.
We use the command \texttt{numactl -i all} for our parallel
experiments.
Our programs use a work-stealing scheduler that we implemented,
implemented similarly to Cilk~\cite{Cilk95}.

\input{inputs/graph_info}
\subsubsection{\nvm{} Configuration}
\label{sec:configurations}
\myparagraph{\nvm{} Modes}
The NVRAM we use (\intelnvm{}) can be configured in two distinct
modes.
In \defn{\memorymode{}}, the DRAM acts like a direct-mapped cache between L3
and the \nvm{} for each socket. \memorymode{} transparently provides
access to higher memory capacity without software modification.  In
this mode, the read-write asymmetry of \nvm{} is obscured by the DRAM
cache, and causes the DRAM hit rate to dominate memory performance.
In \defn{\appdirectmode{}}, \nvm{} acts as byte-addressable storage
independent of DRAM, providing developers with direct access to the
\nvm{}.

\myparagraph{\oursystem{} Configuration}
In \oursystem{}, we configure the \nvm{} to use \appdirectmode{}.  The
devices are configured using the \textsc{fsdax} mode, which removes
the page cache from the I/O path for the device and allows
\textsc{mmap} to directly map to the underlying memory.

\myparagraph{Graph Storage}
The approach we use in \oursystem{} is to store two separate copies of
the graph, one copy on the local \nvm{} of each socket.  Threads can
determine which socket they are running on by reading a thread-local
variable, and access the socket-local copy of the graph.  We discuss
the approach in detail in Section~\ref{sec:graphstorage}

\subsubsection{Graph Data}
To show how our algorithms perform on graphs at different scales, we
selected a representative set of real-world graphs of varying sizes.
These graphs are Web graphs and social networks, which are low-diameter graphs that are
frequently used in practice.
We list the graphs used in our experiments in Table~\ref{table:sizes}, which we symmetrized to obtain larger graphs and so that all of the algorithms would work on them. Hyperlink 2012 is the largest publicly-available real-world graph.
We create weighted graphs for evaluating weighted BFS, Bellman-Ford,
and Widest Path by selecting edge weights in the range $[1, \log n)$
uniformly at random. We process the ClueWeb, Hyperlink2014, and
Hyperlink2012 graphs in the parallel byte-encoded compression format
from Ligra+~\cite{shun2015ligraplus}, and process LiveJournal,
com-Orkut, and Twitter in the uncompressed (CSR) format.

\input{inputs/graphstorage}
\input{inputs/figure_speedup_large}

\input{inputs/our_results}

\input{inputs/figure_us_vs_inmem}

\input{inputs/comparison_to_dram}

\input{inputs/comparison_with_other_nvm}
\input{inputs/comparison_with_semi_external}

%% file: inputs/graph_info.tex
\begin{table}[!t]
\centering
\captionof{table}{\label{table:graphinfo}\small Graph inputs, number
of vertices, edges, and average degree ($\davg$).}
\vspace{0.03in}
\scalebox{0.9}{
\setlength{\tabcolsep}{2pt}
\begin{tabular}[!t]{l|r@{  }|r@{  }|c}
\multicolumn{1}{c|}{Graph Dataset} & \multicolumn{1}{c|}{Num. Vertices} & \multicolumn{1}{c|}{Num. Edges} & $\davg$\\
\midrule
{\emph{LiveJournal}~\cite{boldi2004webgraph} }       & 4,847,571        &85,702,474        &  17.6 \\
{\emph{com-Orkut}~\cite{Yang2015}  }                 & 3,072,627        &234,370,166       &  76.2 \\
{\emph{Twitter}~\cite{kwak2010twitter}  }            & 41,652,231       &2,405,026,092     &  57.7 \\
{\emph{ClueWeb}~\cite{boldi2004webgraph}  }          & 978,408,098      &74,744,358,622    &  76.3 \\
{\emph{Hyperlink2014}~\cite{meusel15hyperlink}  }    & 1,724,573,718    &124,141,874,032   &  72.0 \\
{\emph{Hyperlink2012}~\cite{meusel15hyperlink}  }    & 3,563,602,789    &225,840,663,232   &  63.3 \\
\end{tabular}
}
\label{table:sizes}
\end{table}

%% file: inputs/graphstorage.tex
\subsection{Graph Layout in \nvm{}}\label{sec:graphstorage}

While building \oursystem{}, we observed startingly poor performance of
cross-socket reads to graph data stored on \nvm{}.
We designed a simple micro-benchmark that illustrates this behavior.
The benchmark runs over all vertices in parallel. For the $i$-th
vertex, it counts the number of neighbors incident to it by reducing over
all of its incident edges. It then writes this value to an array
location corresponding to the $i$-th vertex. The graph is stored in
CSR format, and so the benchmark reads each vertex offset exactly
once, and reads the edges incident to each vertex exactly once.
Therefore the total number of reads from the \nvm{} is proportional to
$n + m$, and the number of (in-memory) writes is proportional to $n$.

For the ClueWeb graph, we observed that running the benchmark with the graph
on one socket using all 48 hyper-threads on the \emph{same socket} results
in a running time of 7.1 seconds. However, using \texttt{numactl -i
all}, and running the benchmark on all threads across both sockets
results in a running time of 26.7 seconds, which is \emph{3.7x worse},
despite using twice as many hyper-threads. While we are uncertain as
to the underlying reason for this slowdown, one possible reason could
be the granularity size for the current generation of \nvm{} DIMMs,
which have a larger effective cache line size of 256
bytes~\cite{izraelevitz2019basic}, and a relatively small cache within
the physical NVM device. Using too many threads could cause thrashing,
which is a possible explanation of the slowdowns we observed when
scaling up reads to a single \nvm{} device by increasing the number of
threads. To the best of our knowledge, this significant slowdown has
not been observed before, and understanding how to mitigate it is an
interesting question for future work.

As described earlier, our approach in \oursystem{} is to store two
separate copies of the graph, one on the local \nvm{} of each socket.
Using this configuration, our micro-benchmark runs in
4.3 seconds using all 96 hyper-threads, which is 1.6x faster than the
single-socket experiment and \emph{6.2x faster} than using threads
across both sockets to the graph stored locally within a single socket.

%% file: inputs/figure_speedup_large.tex
\begin{figure*}[!htb]
\begin{center}
\includegraphics[trim={0 0cm 0 0}, width=\textwidth]{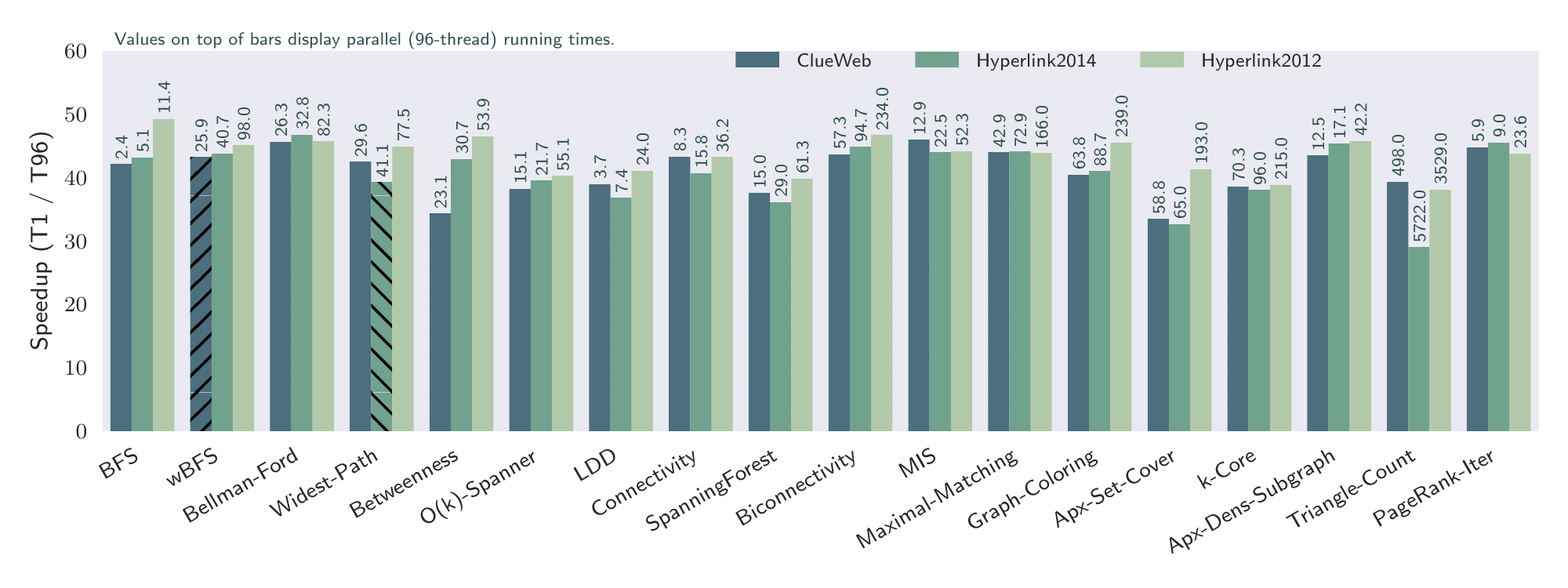}
\caption{\label{fig:speedup_large}
Speedup of \oursystem{} algorithms on large graph inputs on a 48-core
machine (with 2-way hyper-threading), measured relative to the
algorithm's single-thread time. All algorithms are run using NVRAM in
\appdirectmode{}. Each bar is annotated with the parallel running
time on top of the bar.
}
\end{center}
\end{figure*}

%% file: inputs/our_results.tex
\subsection{Scalability}\label{subsec:scalability}
Figure~\ref{fig:speedup_large} shows the speedup obtained on our machine for
\oursystem{} implementations on our large graphs, annotating each bar
with the parallel running time.
In all of these experiments, we store all of the graph data in \nvm{}
and use DRAM for all temporary data.

\myparagraph{Shortest Path Problems}
Our BFS, weighted BFS, Bellman-Ford, and betweenness centrality
implementations achieve between parallel speedups of 31--51x  across all inputs. For $O(k)$-Spanner, we achieve 39--51x speedups
across all inputs. All \oursystem{} codes use the memory-efficient
sparse traversal (i.e., \emapchunk{}) designed in this paper.  We note
that the new weighted-SSSP implementations using \emapchunk{} are up
to 2x more memory-efficient than the implementations
from~\cite{dhulipala18scalable}. We ran our $O(k)$-Spanner
implementation with $k$ set to $\lceil \log_2 n \rceil$ by default.

\myparagraph{Connectivity Problems}
Our low-diameter decomposition implementation achieves a speedup of 28--42x
across all inputs. Our connectivity and spanning forest
implementations, which use the new filtering structure
from Section~\ref{sec:filter},
achieve speedups of 37--53x across all inputs.  Our
biconnectivity implementation
achieves a speed up of 38--46x across
all inputs. We found that setting $\beta=0.2$ in the LDD-based
algorithms (connectivity, spanning forest, and biconnectivity)
performs best in practice, and creates significantly fewer than
$m\beta = m/5$ inter-cluster edges predicted by the theoretical
bound~\cite{miller2013parallel}, due to many duplicate edges that get
removed.

\myparagraph{Covering Problems}
Our MIS, maximal matching, and graph coloring implementations achieve speedups of
43--49x, 33--44x, and 16--39x, respectively.
Our MIS implementation is similar to the implementation from GBBS. Our
maximal matching implementation implements several new optimizations
over the implementation from GBBS, such as using a parallel hash table
to aggregate edges that will be processed in a given round. These
optimizations result in our code (using the graph filter) running
faster than the original code when run in DRAM-only, outperforming the
72-core DRAM-only times reported in~\cite{dhulipala18scalable} for
some graphs (we discuss the speedup of \oursystem{} over GBBS in
Section~\ref{subsec:usvsinmem}).

\myparagraph{Substructure Problems}
Our $k$-core, approximate densest subgraph, and triangle counting
implementations achieve speedups of 9--38x, 43--48x, and
29--63x, respectively.  Our code achieves similar speedups and running
times on \nvm{} compared to the previous times reported
in~\cite{dhulipala18scalable}. We ran the approximate densest subgraph
implementation with $\epsilon=0.001$, which produces subgraphs of
similar density to the $2$-approximation of
Charikar~\cite{charikar00densesubgraph}. Lastly, the \oursystem{}
triangle counting algorithm uses the iterator defined over graph
filters to perform parallel intersection. The performance of our
implementation is affected by the number of edges that must be decoded
for compressed graph inputs, and we discuss this in detail in
Appendix~\ref{apx:filtering}.

\myparagraph{Eigenvector Problems}
Our PageRank implementation achieves a parallel speedup of 42--56x. Our
implementation is based on the PageRank implementation from Ligra, and
improves the parallel scalability of the Ligra-based code by
aggregating the neighbor's contributions for a given vertex in
parallel. We ran our PageRank implementation with $\epsilon = 10^{-6}$
and a damping factor of $0.85$.

%% file: inputs/figure_us_vs_inmem.tex
\begin{figure*}[!htb]
\begin{center}
\includegraphics[width=\textwidth]{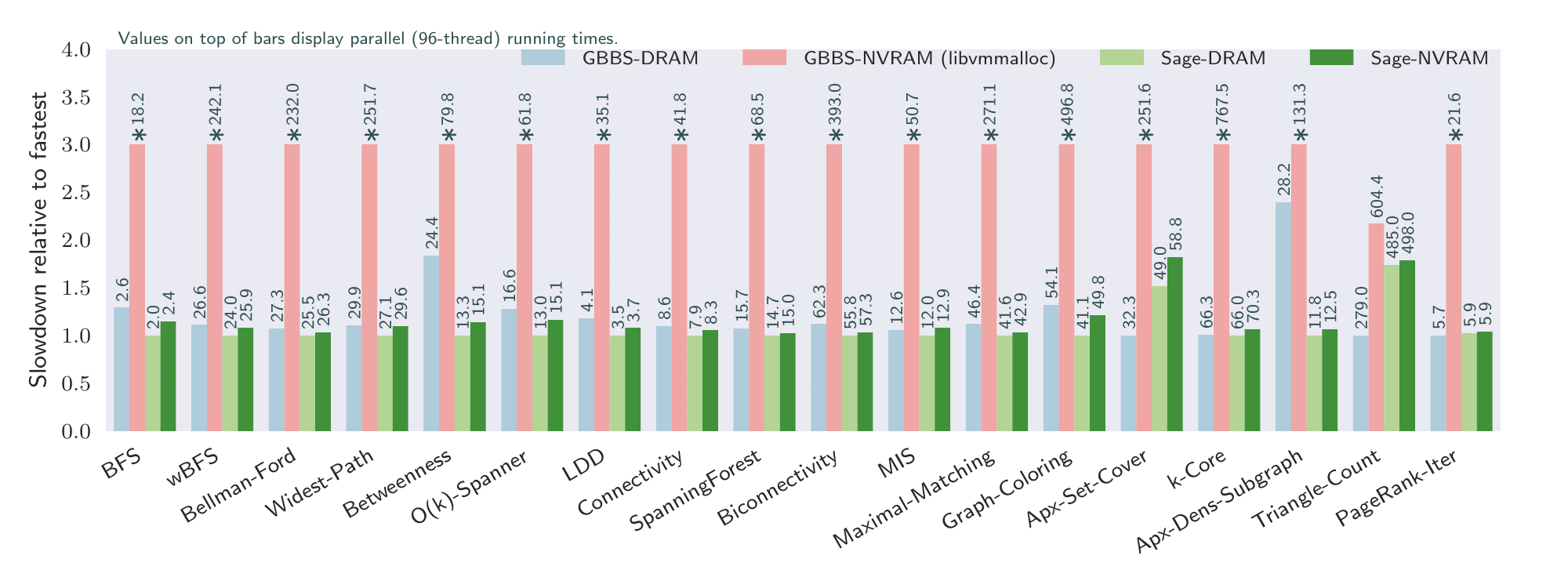}
%\vspace{-1em}
\caption{\label{fig:intro_us_vs_inmem}
Performance of \oursystem{} on the ClueWeb graph compared
with existing state-of-the-art \emph{in-memory} graph processing
systems in terms of slowdown relative to the fastest system (smaller
is better). {GBBS} refers to the DRAM-only codes developed in~\cite{dhulipala18scalable}, and {\oursystem{}} refers to the codes
developed in this paper. Both codes are run in two configurations:
{DRAM} measures the running time when the graph is
stored in memory, and {NVRAM} measures the running time when the graph
is stored in non-volatile memory, and accessed either using the
techniques developed in this paper (\oursystem{}-NVRAM) or using
\textsc{libvmmalloc} to automatically convert the DRAM-only codes from
GBBS to work using non-volatile memory (GBBS-NVRAM). We truncate
relative times slower than 3x and mark the tops of these bars with $*$.
All bars are annotated with the parallel running times of the codes on
a 48-core system with 2-way hyper-threading. Note that the ClueWeb
graph is the largest graph dataset studied in this paper that fits in
the main memory of this machine.
}
\vspace{-1em}
\end{center}
\end{figure*}

%% file: inputs/comparison_to_dram.tex
\subsection{\nvm{} vs. DRAM Performance}\label{subsec:usvsinmem}
In this section, we study how fast \oursystem{} is compared to
state-of-the-art shared-memory graph processing codes, when these
codes are run \emph{entirely in DRAM}.
For these experiments, we study the ClueWeb graph since it is the
largest graph among our inputs where both the graph and all
intermediate algorithm-specific data fully resides in the DRAM of our
machine. We consider the following configurations:
\begin{enumerate}[label=(\textbf{\arabic*}),topsep=1pt,itemsep=0pt,parsep=0pt,leftmargin=15pt]
\item GBBS codes run entirely in DRAM
\item GBBS codes converted to use \nvm{} using libvmmalloc (a robust
      \nvm{} memory allocator)\label{lab:settingtwo}
\item \oursystem{} codes run entirely in DRAM
\item \oursystem{} codes run using \nvm{} in \appdirectmode{}
\end{enumerate}
Setting~\ref{lab:settingtwo} is relevant since it captures the
performance of a naive approach to obtaining \nvm{}-friendly code,
which is to simply run existing shared-memory code using a \nvm{}
memory allocator.

Figure~\ref{fig:intro_us_vs_inmem} displays the results of these
experiments. Comparing \oursystem{} to GBBS when both systems are run
in memory shows that our code is faster than the original GBBS
implementations by 1.17x on average (between 2.38x faster to 1.73x
slower).
The notable exception is for triangle counting, where \oursystem{} is
1.73x slower than the GBBS code (both run in memory). The reason for
this difference is due to the input-ordering the graph is provided in,
and is explained in detail in Appendix~\ref{apx:filtering}.
A number of \oursystem{}
implementations, like connectivity and approximate densest subgraph,
are faster than the GBBS implementations due to optimizations in our
codes that are absent in GBBS, such as a faster implementation of
graph contraction.
Our read-only codes when run using \nvm{} are only about 5\%
slower on average than when run using DRAM-only. This difference in
performance is likely due to the higher cost of \nvm{} reads compared
to DRAM reads. Finally, \oursystem{} is always faster than GBBS when
run on NVRAM using libvmmalloc, and is 6.69x faster on average.

These results show that for a wide range of parallel graph algorithms,
\oursystem{} significantly outperforms a naive approach that converts
DRAM codes to NVRAM ones, is often faster than the fastest DRAM-only
codes when run in DRAM, and is competitive with the fastest DRAM-only
running times when run in NVRAM.

%% file: inputs/comparison_with_other_nvm.tex
\subsection{Alternate \nvm{} approaches}\label{subsec:usvsothernvm}
We now compare \oursystem{} to the fastest available NVRAM approaches
when the input graph \emph{is larger than the DRAM size of the
machine}.
We focus on the Hyperlink2012 graph, which is our only graph where
both the graph and intermediate algorithm data are larger than
DRAM.  We first compare \oursystem{} to the
Galois-based implementations by Gill et al.~\cite{Gill2019}, which use
\nvm{} configured in \memorymode{}.  We then compare \oursystem{} to
the unmodified shared-memory codes from GBBS modified to use \nvm{}
configured in \memorymode{}.

\myparagraph{Comparison with Galois~\cite{Gill2019}}
Gill et al.~\cite{Gill2019} study the performance of several
state-of-the-art graph processing systems, including
Galois~\cite{nguyen2013lightweight}, GBBS~\cite{dhulipala18scalable},
GraphIt~\cite{GraphIt}, and GAP~\cite{BeamerAP15} when run on \nvm{}
configured to use \memorymode{}. Their experiments are run on a nearly
identical machine to ours, with the same amount of DRAM. However,
their machine has 6.144TB of \nvm{} (12 \nvm{} DIMMs with 512GB of
capacity each).

Gill et al.~\cite{Gill2019} find that their Galois-based codes
outperform GAP, GraphIt, and GBBS by between 3.8x, 1.9x, and 1.6x on
average, respectively, for three large graphs inputs, including the
Hyperlink2012 graph. In our experiments running GBBS on \nvm{} using
MemoryMode, we find that the GBBS performance using MemoryMode is 1.3x
slower on average than Galois. There are several possible reasons for
the small difference. First, Gill et al.~\cite{Gill2019} use the
\emph{directed} version of the Hyperlink2012 graph, which has 1.75x
fewer edges than the symmetrized version (225.8B vs.\ 128.7B edges).
The symmetrized graph exhibits a massive connected component
containing 94\% of the vertices, which a graph search algorithm must
process for most source vertices. However, a search from the
largest SCC in the directed graph reaches about half the
vertices~\cite{meusel2014graph}.  Second, they do not enable
compression in GBBS, which is important for reducing the number of
cache-misses and NVRAM reads. Lastly, we found that transparent huge
pages (THP) significantly improves performance, while they did not, which
may be due to differences regarding THP configuration on the
different machines.

Figure~\ref{fig:intro_us_vs_memmode} shows results for
their Galois-based system on the directed Hyperlink2012 graph.
Compared with their \nvm{} codes, \oursystem{} is 1.04--3.08x faster
than their fastest reported times, and 1.94x faster on average. Their
codes use the maximum degree vertex in the directed graph as the
source for BFS, SSSP, and betweenness centrality. We use the maximum
degree vertex in the \emph{symmetric} graph, and note that running on
the symmetric graph is more challenging, since our codes must process
more edges.

Despite the fact that our algorithm must perform more work, our
running times for BFS are 3.08x faster than the time reported
for Galois, and our SSSP time is 1.43x faster. For connectivity
and PageRank, our times are 2.09x faster and 2.12x
faster respectively. For betweenness, our times are 1.04x faster. The
authors also report running times for an implementation of $k$-core
that computes a \emph{single} $k$-core, for a given value of $k$.
This requires significantly fewer rounds than the $k$-core computation
studied in this paper, which computes the coreness number of
\emph{every vertex}, or the largest $k$ such that the vertex
participates in the $k$-core.  They report that their code requires
49.2 seconds to find the $100$-core of the Hyperlink2012 graph. Our
code finds all $k$-cores of this graph in 259 seconds, which requires
running 130,728 iterations of the peeling algorithm and also discovers
the value of the largest $k$-core supported by the graph ($k_{\max} =
10565$).

In summary, we find that using \oursystem{} on \nvm{} using
\appdirectmode{} is 1.94x faster on average than the Galois codes run
using \memorymode{}.

\myparagraph{Algorithms using \memorymode{}}
Next, we compare \oursystem{} to the unmodified shared-memory codes
from GBBS modified to use \nvm{} configured in \memorymode{}. We run
these \memorymode{} experiments on the same machine with 3TB of
\nvm{}, where 1.5TB is configured to be used in \memorymode{}.

Figure~\ref{fig:intro_us_vs_memmode} reports the parallel running
times of both \oursystem{} codes using \nvm{}, and the GBBS codes
using \nvm{} configured in \memorymode{} for the Hyperlink2012 graph.
The results show that in all but one case (triangle counting) our
running times are faster (between 1.15--2.92x). For triangle counting,
the directed version of the Hyperlink2012 graph fits in about 180GB of
memory, which fits within the DRAM of our machine and will therefore
reside in memory.  We note that we also ran \memorymode{} experiments
on the ClueWeb graph, which fits in memory.  The running times were
only 5--10\% slower compared to the DRAM-only running times for the
same GBBS codes reported in Figure~\ref{fig:intro_us_vs_inmem},
indicating a small overhead due to \memorymode{} when the data fits in
memory.

In summary, our results for this experiment show that the techniques
developed in this paper produce meaningful improvements (1.87x speedup
on average, across all 18 problems) over simply running unmodified
shared-memory graph algorithms using \memorymode{} to handle graph
sizes that are larger than DRAM.

%% file: inputs/comparison_with_semi_external.tex
\subsection{External and Semi-External Systems}\label{subsec:usvssemiexternal}
In this section we place \oursystem{}'s performance in context by
comparing it to existing state-of-the-art semi-external memory graph
processing systems.
Table~\ref{table:semiexternalcomp} shows the running times and system
configurations for state-of-the-art results on semi-external memory
graph processing systems. We report the published results presented by
the authors of these systems to give a high-level comparison due to
the fact that (i) our machine does not have parallel SSD devices that
most of these systems require, and (ii) modifying them to use \nvm{}
would be a serious research undertaking in its own right.

\input{inputs/table_semi_external}

\myparagraph{FlashGraph}
FlashGraph~\cite{zheng15flashgraph} is a semi-external memory graph
engine that stores vertex data in memory and stores the edge lists in
an array of SSDs.
Their system is optimized for I/Os at a flash page
granularity (4KB), and merges I/O requests to maximize throughput.
FlashGraph provides a vertex-centric API, and thus cannot implement
some of the work-optimal algorithms designed in \oursystem{}, like our
connectivity, biconnectivity, or parallel set cover algorithms.

We report running times for FlashGraph for Hyperlink2012 on a 32-core
2-way hyper-threaded machine with 512GB of memory and \emph{15 SSDs}
in Table~\ref{table:semiexternalcomp}).  Compared to FlashGraph, the
\oursystem{} times are \emph{9.3x faster on average}. Our BFS and BC
times are 18.2x and 11x faster, and our connectivity, PageRank and
triangle counting implementations are 12.7x, 2.4x faster, and 2.2x
faster, respectively. We note that our times are on the symmetric
version of the Hyperlink2012 graph which has twice the edges, where a
BFS from a random seed hits the massive component containing 95\% of
the vertices (BFSes on the directed graph reach about 30\% of the
vertices).

\hide{
\oursystem{} outperforms FlashGraph from both a cost, engineering, and
algorithmic flexibility perspective, since (i) for good performance
the user must provision an SSD array containing 10s of high-end SSDs
capable of serving close to a million 4KB reads per second (ii)
optimizing for performance involves designing and optimizing locality
in non-standard graph layouts, and (iii) implementations must express
graph programs in the vertex-centric model, compared with the diverse
set of applications that \oursystem{} supports.
}

\myparagraph{Mosaic}
Mosaic~\cite{maass2017mosaic} is a hybrid engine supporting
semi-external memory processing based on a Hilbert-ordered data
structure. Mosaic uses co-processors (Xeon Phis) to offload
edge-centric processing, allowing host processors to perform
vertex-centric operations. Giving a full description of their complex
execution strategy is not possible in this space, but at a high level,
it is based on exploiting the fact that user-programs are written in a
vertex-centric model.

We report the running times for Mosaic run using 1000 hyper-threads,
768GB of RAM, and 6 NVMes in Table~\ref{table:semiexternalcomp}.
Compared with their times, \oursystem{} is 12x faster on average,
solving BFS 1.2x faster, connectivity 44.8x faster, SSSP 3.8x slower,
and 1-iteration of PageRank 2.4x faster.  Given that both SSSP and
PageRank are implemented using an SpMV like algorithm in their system,
we are not sure why their PageRank times are 2.5x slower than the
total time of an SSSP computation. In our experiments, the most costly
iteration of Bellman-Ford takes roughly the same amount of time as a
single PageRank iteration since both algorithms require similar memory
accesses in this step.  \oursystem{} solves a much broader range of
problems compared to Mosaic, and is often faster than
it.

\myparagraph{GridGraph}
GridGraph is an out-of-core graph engine based on a 2-dimensional grid
representation of graphs. Their processing scheme ensures that only a
subset of the vertex-values accessed and written to are in memory at a
given time. GridGraph also offers a mechanism similar to edge
filtering which prevents streaming edges from disk if they are
inactive.
Like FlashGraph, GridGraph is a vertex-centric system and thus cannot
implement algorithms that do not fit in this restricted computational
model.

The authors consider significantly smaller graphs than those used in
our experiments (the largest is a 6.64B edge WebGraph). However, they
do solve the LiveJournal and Twitter graphs that we use.
For the Twitter graph, our BFS and Connectivity times are 15690x
and 359x faster respectively than theirs (our speedups for LiveJournal
are similar). GridGraph does not use direction optimization, which is
likely why their BFS times are much slower.

%% file: inputs/table_semi_external.tex
\setlength{\tabcolsep}{2pt}
\begin{table}[!t]
\centering

\tabcolsep=0.12cm
\captionof{table}{\small
  System configurations (memory in terabytes and threads
  (hyper-threads)) and running times (seconds) of existing
  semi-external memory results on the Hyperlink graphs. The last
  section shows our running times (note that our system is also
  equipped with \nvm{} DIMMs). *These problems are run on directed
  versions of the graph.
}
\scalebox{0.85}{
\begin{tabular}[!t]{lllrrrr}
\toprule
Paper & Problem & Graph & Mem & Threads & Time \\
\midrule

\multirow{5}{*}{FlashGraph~\cite{zheng15flashgraph}}
& BFS*          & 2012 & .512 & 64 & 208  \\
& BC*           & 2012 & .512 & 64 & 595  \\
& Connectivity* & 2012 & .512 & 64 & 461  \\
& PageRank*     & 2012 & .512 & 64 & 2041 \\
& TC*           & 2012 & .512 & 64 & 7818 \\
\midrule %9.3x avg

\multirow{4}{*}{Mosaic~\cite{maass2017mosaic}}
& BFS*                     & 2014 & 0.768  & 1000 &  6.55  \\
& Connectivity*            & 2014 & 0.768  & 1000 &  708   \\
& PageRank (1 iter.)*      & 2014 & 0.768  & 1000 &  21.6 \\ 
& SSSP*                    & 2014 & 0.768  & 1000 &  8.6   \\
\midrule % 12.1x avg

\multirow{8}{*}{\oursystem{}}
& BFS                 & 2014 & 0.375 & 96 & 5.10 \\
& SSSP                & 2014 & 0.375 & 96 & 32.8 \\
& Connectivity        & 2014 & 0.375 & 96 & 15.8 \\
& PageRank (1 iter.)  & 2014 & 0.375 & 96 & 8.99 \\
& BFS                 & 2012 & 0.375 & 96 & 11.4 \\
& BC                  & 2012 & 0.375 & 96 & 53.9 \\
& Connectivity        & 2012 & 0.375 & 96 & 36.2 \\
& SSSP                & 2012 & 0.375 & 96 & 82.3 \\
& PageRank            & 2012 & 0.375 & 96 & 827  \\
& TC                  & 2012 & 0.375 & 96 & 3529 \\
\bottomrule

\end{tabular}
}
\label{table:semiexternalcomp}
\end{table}

%% file: inputs/related.tex
\section{Related Work}\label{sec:related}

A significant amount of research has focused on reducing
expensive writes to NVRAMs.  Early work has designed algorithms for
database operators~\cite{Chen11,Viglas12,viglas2014write}.
Blelloch et al.~\cite{BBFGGMS16,BFGGS15,blelloch2016efficient}
define computational models to capture the asymmetric
read-write cost on NVRAMs, and many algorithms and lower bounds have been
obtained based on the
models~\cite{BBFGGMS18implicit, blelloch2020improved, blelloch2018parallel, Gu2018, jacob2017}.
Other models and systems to reduce writes or memory footprint on
NVRAMs have also been described~\cite{arulraj2018bztree, ArulrajP17,
cai2019matrix, carson2016write, chen2019design, lee2017wort,
liu2019write, nissim2019revisiting, oukid2017memory,
shen2017efficient, van2018managing}.

Persistence is a key property of NVRAMs due to their non-volatility.
Many new persistent data structures have been designed for
NVRAMs~\cite{attiya2019tracking, ben2019delay, chen2015persistent,
  cohen2018inherent, panhart, shull2019autopersist}.
There has also been research on automatic recovery schemes and transactional memory for NVRAMs~\cite{alshboul2019efficient, correia2018romulus,
lersch2019persistent, liu2018ido, wang2019crash,
zhang2019pangolin, zhou2019brief}.
There are several recent papers benchmarking performance on
NVRAMs~\cite{izraelevitz2019basic, liu2019initial, van2019persistent}.

Parallel graph processing frameworks have received significant
attention due to the need to quickly analyze large
graphs~\cite{sahu2017ubiquity}. The only previous graph processing
work targeting NVRAMs is the concurrent work by Gill et
al.~\cite{Gill2019}, which we discuss in Section~\ref{subsec:usvsothernvm}.
Dhulipala et al.~\cite{dhulipala2017julienne,dhulipala18scalable}
design the Graph Based Benchmark Suite, and show that the largest
publicly-available graph, the Hyperlink2012 graph, can be efficiently
processed on a single multicore machine. We compare with these
algorithms in Section~\ref{sec:exps}. Other multicore frameworks
include Galois~\cite{nguyen2013lightweight},
Ligra~\cite{ShunB2013,shun2015ligraplus},
Polymer~\cite{zhang2015polymer}, Gemini~\cite{Gemini},
GraphGrind~\cite{Sun2017}, Green-Marl~\cite{Hong2012},
Grazelle~\cite{Grossman2018}, and
GraphIt~\cite{GraphIt}. We refer the reader
to~\cite{ammar2018experimental, McCune2015, shi2018, Yan2017} for
excellent surveys of this growing literature.

%% file: inputs/conclusion.tex
\section{Conclusion}
We have introduced \oursystem{}, which takes a semi-asymmetric
approach to designing parallel graph algorithms that avoid writing to
the \nvm{} and uses DRAM proportional to the number of vertices. We
have designed a new model, the \SAM{}, and have shown that all of our
algorithms in \oursystem{} are provably efficient, and often
work-optimal in the model. Our empirical study shows that \oursystem{}
graph algorithms can bridge the performance gap between NVRAM and
DRAM. This enables NVRAMs, which are more cost-efficient and support
larger capacities than traditional DRAM, to be used for large-scale
graph processing.
Interesting directions for future work include studying which filtering
algorithms can be made to use only $O(n)$ words of DRAM, and to study
how \oursystem{} performs relative to existing NVRAM graph-processing
approaches on synthetic graphs with trillions of edges.

%% file: inputs/appendix.tex
\begin{appendix}
  \input{inputs/blocking-apx}

  \input{inputs/bucketing-apx}

  \input{inputs/algorithm-specifications}
  \input{inputs/experiments-apx}

\end{appendix}

%% file: inputs/blocking-apx.tex
\section{Memory-Efficient Parallel Graph Traversal}\label{apx:blocking}

In this section, we provide additional details on \oursystem{}'s
memory-efficient parallel graph traversal algorithm. The algorithm
description uses the parallel primitives defined in
Section~\ref{sec:prelims}.

\myparagraph{Chunked Parallel Traversal}
In \oursystem{}, we use a chunk-based method to optimize the
memory-efficiency of the sparse (push-based) \emap{}. Our approach,
which we refer to as \emapchunk{}, shares some similarities, and
achieves the same cache performance as the \emapblock{} implementation
used in GBBS~\cite{dhulipala18scalable}, but crucially it
improves the intermediate memory usage.

We provide pseudocode for our algorithm in
Algorithm~\ref{alg:emapchunk}. The algorithm is based on three types
of chunking.
First, the algorithm breaks up the edges incident to each vertex into
\emph{blocks} based on the underlying graph's block size.
Second, the algorithm chunks the outgoing edges to traverse, which it
breaks up into \emph{groups}. Each group consists of some number of
blocks. The blocks within a group will be processed sequentially, but
different groups can be processed in parallel.
Third, the algorithm performs \emph{chunking} of the output that is
generated, writing out the neighbors that must be emitted in the
output \vset{} into fixed-size \emph{chunks}.

The algorithm first breaks each vertex up into units of work called
blocks based on a block size parameter, $G_{\codevar{b\_size}}$
(Line~\ref{line:bs}). This block size can be tuned arbitrarily for
uncompressed graphs, but must be set equal to the compression block
size for compressed graphs.  Different settings of the block size
result in a tradeoff between the depth of the algorithm and the amount
of \smallmem{} used in the \SAMabbrev{} model. We discuss this
tradeoff, and how to set this block size below.
Next, the algorithm decides the number of groups to create
(Lines~\ref{line:prefixsum}--\ref{line:search}), where each group
consists of a set of blocks. It then processes the groups in parallel.
For each group, it processes the blocks within the group one at a
time. When starting the next block, it calls the \textsc{FetchChunk}
procedure (Lines~\ref{line:fcstart}--\ref{line:fcend}), which returns
an output chunk for the current group, allocating a fresh chunk if the
current chunk is too full (Line~\ref{line:fetchchunk}). Each group
stores the chunks allocated for it in a per-group vector of output
chunks ($\codevar{V}$ from Line~\ref{line:vectorofchunks}), which can
be accessed safely without any atomics, since each group is processed
by a single thread. The chunk allocations are done in our
implementation using a pool-based thread-local allocator
(Line~\ref{line:chunkalloc}). Next, the algorithm processes the block
and writes all neighbors that should be emitted in the next \vset{}
into the chunk, and updates the block size
(Line~\ref{line:processblock}). Note that the \textsc{FetchChunk}
procedure ensures that the returned chunk has sufficient space to
store all neighbors in the block being processed
(Line~\ref{line:fccheck}).  The remaining steps aggregate the chunks
from the per-group vectors (Line~\ref{line:allchunks}), perform prefix
sum on the chunk sizes (Line~\ref{line:chunkprefixsum}), and copy the
data within the chunks into an array with size proportional to the
number of returned neighbors
(Lines~\ref{line:copystart}--\ref{line:copyandrelease}).  After
copying the data within a chunk, the algorithm frees the chunk
(Line~\ref{line:copyandrelease}).  Finally, the algorithm returns the
output \vset{} (Line~\ref{line:output}).

\myparagraph{Memory Usage, Work, and Depth}
First note that in the degenerate case where \emph{all} edges are
processed using our implementation, the code can create up to
$m/G_{\codevar{b\_size}}$ many blocks, which can be $\Omega(n)$.
Instead, we ensure that $G_{\codevar{b\_size}} = d_{\emph{avg}} =
\lceil m/n\rceil $, or the average degree.  In this case, the maximum
number of blocks used is $m/G_{\codevar{b\_size}} = m/d_{\emph{avg}} =
O(n)$. It is simple to check the remainder of the code and observe
that the amount of intermediate memory and the output size are
bounded by $O(n)$ words. The overall \smallmem{} usage of the
procedure is therefore $O(n)$ words. Note that for compressed graphs,
the block size is equal to the compression block size, and thus the
compression block size must be set to $d_{\emph{avg}}$ to ensure
theoretical efficiency in the \SAMabbrev{}. For uncompressed graphs
the block size can be set arbitrarily.

To ensure that we do not create an unnecessarily large number of
groups, we set the number of groups to
$O(\min(8p,\allowbreak\sum_{u\in U}\emph{deg}(u)/\codevar{min\_group\_size}))$ on $p$
processors
(Lines~\ref{line:groupsizestart}--\ref{line:groupsizeend} and
Line~\ref{line:mingroupsize}). These parameters balance between
providing enough parallel slackness for work-stealing when there is a
large amount of work to be done (the $8p$ term in the $\min$) while
also ensuring that we do not over-provision parallelism in the case
when the number of edges incident to the frontier is small (the second
term in the $\min$).

The overall work of the procedure is $O(\sum_{u\in U}\emph{deg}(u))$,
and the overall depth is $O(\log n + \codevar{group\_size})$, since each
group is processed sequentially by a single thread, and each call to
the procedure requires aggregating the per-group chunks, which can be
done in parallel. Note that theoretically, $\codevar{group\_size} =
O(d_{\emph{avg}})$ since the $\max$ on Line~\ref{line:setgroupsize}
includes the $\lceil d_{U} / 8p \rceil$ only to avoid creating excess
groups when parallelism is abundant, and  $\codevar{min\_group\_size}
= O(d_{\emph{avg}})$ (Line~\ref{line:mingroupsize}).  The work done
for the chunk-aggregation is proportional to the number of groups (and
number of output chunks), both of which are upper-bounded by
$O(\sum_{u\in U}\emph{deg}(u))$.  $d_{\emph{avg}}$ is usually a small
constant in real-world graphs (see Table~\ref{table:sizes}).

\algblock{ParFor}{EndParFor}
\algnewcommand\algorithmicparfor{\textbf{parfor}}
\algnewcommand\algorithmicpardo{\textbf{do}}
\algnewcommand\algorithmicendparfor{}
\algrenewtext{ParFor}[1]{\algorithmicparfor\ #1\ \algorithmicpardo}
\algrenewtext{EndParFor}{\algorithmicendparfor}
\algtext*{EndParFor}{}

\algrenewcommand\algorithmicindent{1.0em}%
\begin{algorithm}[!t]
\caption{\emapchunk{}}\label{alg:emapchunk}
\small
\begin{algorithmic}[1]
\State $\codevar{chunk\_size} = \max(4096, d_{\emph{avg}})$
\State $\codevar{min\_group\_size} = \max(4096, d_{\emph{avg}})$\label{line:mingroupsize}
\State $\codevar{chunk\_alloc} \coloneqq$ Initialize thread-local allocator with $\codevar{chunk\_size}$\label{line:chunkalloc}
\Procedure{FetchChunk}{$\codevar{b\_size}$, $V$}\label{line:fcstart}
  \State $\codevar{chunk} \coloneqq $ Last chunk from $V$
    \Comment{$V : $ vector of output chunks for this group}
  \If{$\codevar{chunk} = \mathsf{null}$ or $\codevar{chunk}.\textsc{size} + b\_size > \codevar{chunk\_size}$}\label{line:fccheck}
    \State $\codevar{chunk} \coloneqq \codevar{chunk\_alloc}.\textsc{allocate}()$
    \State Insert $\codevar{chunk}$ into $V$
  \EndIf
  \State \algorithmicreturn{} $\codevar{chunk}$\label{line:fcend}
\EndProcedure

\Procedure{\emapchunk{}}{$G, U, F$}
\State $G_{\codevar{b\_size}} \coloneqq d_{\emph{avg}}$ \Comment{underlying block size used in $G$}\label{line:bs}
\State $\codevar{B} \coloneqq$ Output blocks corresponding to $u \in U$\label{line:outblocks}
\State $\codevar{O} \coloneqq$ Prefix sums of block-degrees for blocks in $B$\label{line:prefixsum}
\State $d_{U} \coloneqq \sum_{u \in U} \emph{deg}(u)$\label{line:groupsizestart}
\State $\codevar{group\_size} \coloneqq \max(\lceil d_{U} / 8p \rceil, \codevar{min\_group\_size})$\label{line:setgroupsize}
\State $\codevar{num\_groups} \coloneqq \lceil d_{U} / \codevar{group\_size} \rceil$\label{line:groupsizeend}
\State $\codevar{idxs} \coloneqq \{i\cdot \codevar{group\_size} \ | \ i \in [\codevar{num\_groups}]\}$
\State $\codevar{Offs} \coloneqq$ Offsets into $O$ resulting from a parallel merge of $idxs$ and $O$ \label{line:search}
\State $\codevar{V} \coloneqq $ Array of vectors storing chunks of size $\codevar{num\_groups}$\label{line:vectorofchunks}
\ParFor {$i$ in $[0, |\codevar{Offs}|)$} \Comment{In parallel over the groups}
  \For {$j$ in $[\codevar{Offs}[i], \codevar{Offs}[i+1])$} \Comment{$\leq G_{\codevar{b\_size}}$ edges in block $j$}
    \State $\codevar{chunk} \coloneqq
    \textproc{FetchChunk}(G_{\codevar{b\_size}}, V_i)$\label{line:fetchchunk}
    \State Process vertices in block $j$ by applying $F$, and write
    vertices where $F$ returns $\mathsf{true}$ into $\codevar{chunk}$\label{line:processblock}
  \EndFor
\EndParFor
\State $\codevar{C} \coloneqq$ All chunks extracted from $\codevar{V}$\label{line:allchunks}
\State $\codevar{output\_offsets} \coloneqq$ Array of length $|C|$
where the $i$'th entry contains the size of the $i$'th chunk, $C[i]$
\State $\codevar{output\_size} \coloneqq$ $\mathit{Scan}(\codevar{output\_offsets}, +)$\label{line:chunkprefixsum}
\State $\codevar{output} \coloneqq$ Array of size $\codevar{output\_size}$\label{line:outputarray}
\ParFor {$c \in \codevar{C}$}\label{line:copystart}
  \State Copy elements in $c$ to $\codevar{output}$ starting at offset in $\codevar{output\_offsets}$ corresponding to $c$
	\State $\codevar{chunk\_alloc}.\textsc{release}(c)$\label{line:copyandrelease}
\EndParFor
\State \algorithmicreturn{} $\codevar{output}$\label{line:output}
\EndProcedure
\end{algorithmic}
\end{algorithm}

%% file: inputs/bucketing-apx.tex
\section{Semi-Asymmetric Bucketing}\label{apx:bucketing}
We now briefly describe how to adapt the work-efficient bucketing
structure from Julienne~\cite{dhulipala2017julienne} to the
\SAMabbrev{}.  A bucketing structure maintains a dynamic mapping
between a set of \emph{elements} and \emph{buckets}, and is used in
several important graph algorithms for work-efficiency: weighted
breadth-first search, $k$-core, approximate densest subgraph, and
approximate set cover.  The bucketing strategy in Julienne is based on
\emph{lazy bucketing}, which avoids deleting the bucketed elements
from buckets that they are moved out of. If the elements that are bucketed
are the vertices, and the total number of bucket updates is $O(m)$, then
the use of lazy bucketing will require the bucket structure to use
$O(m)$ words of small-memory, violating the \SAMabbrev{} requirements.

We can address this issue by using \emph{semi-eager bucketing}. In the
semi-eager version, each bucket maintains two counters, storing the
number of live (not logically deleted) elements currently in the
bucket, and the number of dead (logically deleted) elements. When
moving a vertex out of a bucket, we increment the dead element count
in that bucket. When a bucket contains more than a constant factor of
dead elements, we physically pack them out. Since each vertex is
contained in a single bucket, this approach only uses $O(n)$ words of
small-memory for bucketing vertices.

In practice, we use the practical variant of the bucketing structure
proposed in Julienne~\cite{dhulipala2017julienne}, which is based on
maintaining a constant number of active buckets with the highest
priority, and an overflow bucket for all remaining vertices.
This approach also uses only $O(n)$ words of small-memory for
bucketing vertices.

%% file: inputs/algorithm-specifications.tex
\section{Algorithm Specification and
Analysis}\label{apx:algorithmspecs}

This section supplements Section~\ref{sec:algorithms} with
specifications of the problems that we study, and additional algorithm
details.

\subsection{Shortest Path Problems}\label{sec:sssp}
\begin{benchmark}{Breadth-First Search (BFS)}
Input: $G=(V, E)$, an unweighted graph, $\src \in V$.\\
Output: $P$, a mapping containing parent-pointers for a BFS tree
rooted at $\src$. Specifically, for $v \in
V$ reachable from $\src$ and not equal to $\src$, $P[v] = p_v$ where
$\dist_{G}(\src, p_v) + 1 = \dist_{G}(\src, v)$, $P[\src] = \src$, and
$P[v] = \infty$ if $v$ is unreachable from $\src$. $\dist_{G}(u,v)$ is
the shortest path distance between $u$ and $v$ in $G$.
\end{benchmark}
\begin{benchmark}{Integral-Weight SSSP (Weighted BFS)}
Input: $G=(V, E, w)$, a weighted graph with integral edge weights, $\src \in V$.\\
Output: $D$, a mapping where $D[v]$ is the shortest path distance from $\src$ to $v$ in $G$ and $\infty$ if $v$ is unreachable.
\end{benchmark}
\begin{benchmark}{General-Weight SSSP (Bellman-Ford)}
Input: $G=(V, E, w)$, a weighted graph, $\src \in V$.\\
Output: $D$, a mapping where $D[v]$ is the shortest path distance from $\src$ to $v$ in $G$ and $\infty$ if $v$ is unreachable.
If the graph contains any negative-weight cycles reachable from
$\src$, the vertices of these negative-weight cycles and vertices
reachable from them must have a distance of $-\infty$.
\end{benchmark}
\begin{benchmark}{Single-Source Betweenness Centrality (BC)}
Input: $G=(V, E)$, an undirected graph, $\src \in V$.\\
Output: $S$, a mapping from each vertex $v$ to the betweenness centrality contribution from all $(\src, t)$ shortest paths that pass through $v$.
\end{benchmark}
\begin{benchmark}{Integral-Weight Single-Source Widest Path}
Input: $G=(V, E, w)$, a weighted graph with integral edge weights, $\src \in V$.\\
Output: $D$, a mapping where $D[v]$ is the maximum over all paths
between $\src$ and $v$ in $G$ of the minimum weight on the path, and
$-\infty$ if $v$ is unreachable.
\end{benchmark}
\begin{benchmark}{$O(k)$-Spanner}
Input: $G=(V, E)$, an undirected, unweighted graph, and an integer
stretch factor, $k$.\\
Output: $H \subseteq E$, a set of edges such that for every pair of vertices $u,v \in V$ that are
connected in $G$, $\dist_{H}(u, v) \leq O(k) \cdot \dist_{G}(u,v)$,
where $\dist_{G}(u,v)$ ($\dist_{H}(u,v)$) is the shortest path distance
between $u$ and $v$ in $G$ ($H$, respectively).
\end{benchmark}

We consider six shortest-path problems in this paper: breadth-first
search (BFS), integral-weight SSSP (wBFS), general-weight SSSP
(Bellman-Ford), single-source betweenness centrality, integral-weight single-source
widest path (widest-path), and $O(k)$-spanner. Our implementations of BFS,
Bellman-Ford, and betweenness centrality are based on the
implementations in Ligra~\cite{ShunB2013}, and our wBFS
implementation is based on our earlier work on
Julienne~\cite{dhulipala2017julienne}. We provide two implementations
of the widest-path algorithm, one based on Bellman-Ford, and another
based on the wBFS implementation from
Julienne~\cite{dhulipala2017julienne}. As mentioned in
Section~\ref{sec:shortestpath}, we obtain semi-asymmetric algorithms
for the first five these problems by using  \emapchunk{}
(Section~\ref{subsec:blocking}) for performing sparse graph
traversals.

Our $O(k)$-spanner implementation is based on a recently described
spanner algorithm due to Miller et al.~\cite{miller2015spanners}. A $O(k)$-spanner is a subgraph that
preserves shortest path distances within a multiplicative factor of
$O(k)$. Their construction computes an $O(k)$-spanner with size
$O(n^{1+1/k})$, and runs in $O(m)$ expected work and $O(k\log n)$ depth \whp{}. The
implementation uses a low-diameter decomposition algorithm, which we
describe below. In our experiments, we set $k$ to be
$\lceil \log n \rceil$, which gives a spanner that has $O(n)$ size.

\subsection{Connectivity Problems}\label{sec:connectivityproblems}
\begin{benchmark}{Low-Diameter Decomposition (LDD)}
{Input:} $G=(V, E)$, a directed graph, $0 < \beta < 1$.\\
{Output:} $\mathcal{L}$, a mapping from each vertex to a cluster ID representing a $(O(\beta), O((\log n)/\beta))$ decomposition.
A $(\beta, d)$-decomposition partitions $V$ into $V_{1}, \ldots, V_{k}$ such that
the shortest path between two vertices in $V_{i}$ using only vertices in
$V_{i}$ is at most $d$, and the number of edges $(u,v)$ where
$u \in V_{i}, v \in V_{j}, j \neq i$ is at most $\beta m$.
\end{benchmark}
\begin{benchmark}{Connectivity}
Input: $G=(V,E)$, an undirected graph.\\
Output: $\mathcal{L}$, a mapping from each vertex to a unique label for its connected component.
\end{benchmark}
\begin{benchmark}{Spanning Forest}
Input: $G=(V,E)$, an undirected graph.\\
Output: $T$, a set of edges representing a spanning forest of $G$.
\end{benchmark}
\begin{benchmark}{Biconnectivity}
Input: $G=(V,E)$, an undirected graph.\\
Output: $\mathcal{L}$, a mapping from each edge to the label of its biconnected component.
\end{benchmark}

We consider four connectivity
problems in this paper: low-diameter decomposition (LDD),
connectivity, spanning forest, and biconnectivity. Our implementations
in this paper are extensions of the implementations provided in
GBBS~\cite{dhulipala18scalable}. We note that connectivity, spanning
forest, biconnectivity, and $O(k)$-spanner all use LDD as a
subroutine.
Our results for connectivity, spanning forest, and biconnectivity work
by observing that the low-diameter decomposition primitive used in the
work-efficient connectivity algorithm of Shun et
al.~\cite{shun2014practical} produces a much larger drop in the number
of inter-cluster edges than the original analysis of this algorithm.
In particular, the number of inter-cluster edges becomes $O(n)$ in
expectation after a single application of LDD. Therefore, we can
simply run a single round of LDD, and build the inter-cluster graph
in the fast memory after a single round. The specific result we use is
Corollary 3.1 of ~\cite{miller2015spanners}, which we reproduce here
for completeness:

\begin{lemma}[Corollary 3.1
of~\cite{miller2015spanners}]\label{lem:millerpengxuvladu}
In an exponential start time decomposition with parameter $\beta =
\frac{\log n}{2k}$, for any vertex $v \in V$ the ball
$B(v,1) = \{u \in V\ |\ d(u,v) \leq 1\}$ intersects $O(n^{1/k})$ clusters in
expectation.
\end{lemma}

The number of clusters the radius-1 ball around a vertex intersects is
exactly the number of inter-cluster edges incident to the vertex.
Applying Lemma~\ref{lem:millerpengxuvladu} with $k = c\log n$, we have
that $\beta = \frac{1}{2c}$, for a suitable constant $c$, and that the
number of inter-cluster edges incident to $v$ is $O(n^{1/c\log n}) =
O(1)$ in expectation.

Thus, applying this lemma across all vertices, by linearity the number
of inter-cluster edges in $G$ drops to $O(n)$ in expectation after
applying the LDD once, when $\beta = \frac{1}{2c}$ for some constant
$c$. At this point, the entire graph on $n$ vertices, and all
inter-cluster edges can be built in the fast memory, and the previous
algorithm of Shun et al.~\cite{shun2014practical} can be run in the
fast memory on this new graph. By the analysis above, and the fact
that the Shun et al.~\cite{shun2014practical} algorithm runs in $O(m)$
expected work and $O(\log^{3} n)$ depth \whp{}, we have the following
result for connectivity in the \SAMabbrev{}.

\begin{theorem}\label{thm:conn}
There is a parallel connectivity algorithm that performs $O(m)$
expected work and $O(\log^{3} n)$ depth \whp{} in the \SAMabbrev{}
model with $O(n)$ words of memory.
\end{theorem}

An identical argument holds for spanning forest, and biconnectivity
(which uses the connectivity algorithm as a subroutine).

\begin{corollary}\label{cor:fastsfandbicc}
There are algorithms for computing spanning forest and
biconnectivity that perform $O(m)$ expected work each, and $O(\log^{3}
n)$ depth and $O(d_{G}\log n + \log^{3} n)$ depth \whp{} in the
\SAMabbrev{} model with $O(n)$ words of memory.
\end{corollary}

We note that our biconnectivity implementation utilizes the filtering
structure (Section~\ref{sec:filter}) to accelerate a call to
connectivity that operates on a graph with most of the edges removed,
and thus our implementation runs in the relaxed \SAMabbrev{} model for
practical efficiency.

\myparagraph{Handling Restarts in Low-Diameter Decomposition-Based Algorithms}
One issue that arises when running in the \SAMabbrev{} is that
algorithms which use $O(n)$ words of small-memory \emph{in
expectation} may need to be restarted if the space bound is violated.
This situation can occur for several of our algorithms which are based
on running a low-diameter decomposition, and contracting, or selecting
inter-cluster edges based on the decomposition. In what follows, we
focus on the connectivity algorithm, since our spanning forest,
biconnectivity, and $O(k)$-spanner algorithm are also handled
identically.

In our connectivity algorithm, we can ensure that the algorithm runs
in $O(n)$ words of space by re-running the LDD algorithm until it
succeeds.  Checking whether an LDD succeeds can be done in $O(n)$
space and $O(n+m)$ work by simply counting the number of inter-cluster
edges formed by the partition of vertices.  Once the LDD succeeds, the
rest of the algorithm does not require restarts, since the recursive
calls in the connectivity algorithm always fit within $O(n)$ words of
space~\cite{shun2014practical}. We note that since the LDD succeeds
with constant probability, an expected constant number of iterations
are needed for the connectivity algorithm to succeed.  Therefore, the
work bounds obtained using restarting are still $O(m)$ in
expectation. Furthermore, the depth bound is not affected by the
restarts since we only need to perform restarts at the first level of
recursion in the algorithm. Since we must perform at most $O(\log n)$
restarts at this level to guarantee success \whp{}, the overall
contribution to the depth of re-running an LDD at this step is
$O(\log^2 n)$ \whp{}, which is subsumed by the algorithm's overall
depth.

\subsection{Covering Problems}\label{sec:coveringproblems}
\begin{benchmark}{Maximal Independent Set (MIS)}
Input: $G=(V, E)$, an undirected graph.\\
Output: $U \subseteq V$, a set of vertices such that no two vertices in $U$ are neighbors and all vertices in $V \setminus U$ have a neighbor in $U$.
\end{benchmark}
\begin{benchmark}{Maximal Matching}
Input: $G=(V, E)$, an undirected graph.\\
Output: $E' \subseteq E$, a set of edges such that no two edges in $E'$ share an endpoint and all edges in $E \setminus E'$ share an endpoint with some edge in $E'$.
\end{benchmark}
\begin{benchmark}{Graph Coloring}
Input: $G=(V, E)$, an undirected graph.\\
Output: $C$, a mapping from each vertex to a color such that for each
edge $(u, v) \in E$, $C(u) \neq C(v)$, using at most $\Delta+1$ colors.
\end{benchmark}
\begin{benchmark}{Approximate Set Cover}
Input: $G=(V, E)$, an undirected graph representing a set cover instance.\\
Output: $S \subseteq V$, a set of sets such that $\cup_{s \in s} N(s) = V$ with
$|S|$ being an $O(\log n)$-approximation to the optimal cover.
\end{benchmark}

We consider four covering problems in this paper: maximal independent
set (MIS), maximal matching, graph coloring, and approximate set
cover. Our implementations are extensions of the implementations
described in~\cite{dhulipala18scalable}. First, for
MIS and graph coloring, we derive \SAMabbrev{} algorithms by applying
our \emapchunk{} optimization since other than graph traversals, both
algorithms already use $O(n)$ words of \smallmem{}. Both maximal matching and
approximate set cover require using our graph filtering technique to
achieve immutability and reduced memory usage without affecting the
theoretical bounds of the algorithms.

Our maximal matching algorithm runs phases of the filtering-based
maximal matching algorithm described in~\cite{dhulipala18scalable} on
subsets of the edges such that they fit in \smallmem{}.  The algorithm
has access to $O(n + m/\log n)$ words of memory, so we can solve the
problem by extracting the next $O(m/\log n)$ unprocessed edges on each
phase, and running the random-priority based maximal matching
algorithm on the subset of edges. The algorithm will finish after
performing $O(\log n)$ phases. This does not affect the
overall work and increases the depth by an $O(\log n)$ factor. The
depth of applying maximal matching within a phase is $O(\log^2 n)$
\whp{} using the analysis of Fischer and Noever~\cite{FischerN19},
combined with the implementation from Blelloch et
al.~\cite{blelloch2012greedy}. Thus, the overall depth of the
algorithm is $O(\log^3 n)$ \whp{}.
In practice, we use a similar approach that is theoretically motivated
and makes use of our graph-filtering structure. In each phase, we
extract $O(n)$ unmatched edges, and process them using the
random-priority based algorithm~\cite{blelloch2012greedy}. All
unmatched edges from this set are discarded, and the graph is filtered
using \filteredges{} to pack out edges incident to matched vertices.
Theoretically, we can switch to the previously described version after
a constant number of such phases.  However, we observed that in
practice, a constant number of iterations of the filtering procedure
suffices for all graphs that we tested on.

Our approximate set cover implementation is similar to the
implementation from~\cite{dhulipala18scalable}, with the exception
that the underlying filtering is done using a graph filter, instead of
mutating the original graph. The bounds obtained for the graph
filter structure match the bounds on filtering used in the GBBS code,
which mutates the underlying graph, and so our implementation also
computes a $(1+\epsilon)$-approximate set cover in $O(m)$ expected
work and $O(\log^{3} n)$ depth \whp{}.

\subsection{Substructure Problems}\label{sec:substructureproblems}
\begin{benchmark}{$k$-core}
Input: $G=(V, E)$, an undirected graph.\\
Output: $D$, a mapping from each vertex to its coreness value, i.e., the maximum $k$-core a vertex participates in.
The $k$-core of $G$ is a maximal set of vertices, $C$, such that each $v
\in C$ has degree at least $k$ in $G[C]$, the induced subgraph of $C$ in $G$.
\end{benchmark}
\begin{benchmark}{Approximate Densest Subgraph}
Input: $G=(V, E)$, an undirected graph, and a parameter $\epsilon$.\\
Output: $U \subseteq V$, a set of vertices such that the density of $G_{U}$
is a $2(1+\epsilon)$ approximation of density of the densest subgraph
  of $G$.\\
Definition: Given an undirected graph $G=(V, E)$, the density of a set $S
\subseteq V$ is defined as $\rho(S) = |E(G[S])|/|S|$, where $E(G[S])$
are the edges in the induced subgraph on $S$.
\end{benchmark}
\begin{benchmark}{Triangle Counting}
Input: $G=(V, E)$, an undirected graph.\\
Output: The total number of triangles in $G$.
\end{benchmark}

We consider three substructure problems in this paper: $k$-core,
approximate densest subgraph and triangle counting. Our $k$-core
implementation is similar to the implementation from
GBBS~\cite{dhulipala18scalable}. We refer to
Section~\ref{sec:substructure} for descriptions of our $k$-core and
approximate densest subgraph algorithms.

Our triangle counting code is based on the GBBS implementation, which
is an implementation of the parallel triangle counting algorithm from
Shun and Tangwongsan~\cite{shun2015multicore}. Our implementation uses
the graph filter structure to orient edges in the graph from lower
degree to higher degree. Since we only require outgoing edges, we
supply a flag to the framework which permits it to only represent the
outgoing edges of the directed graph, halving the amount of internal
memory required. Our implementation uses the new iterator implemented
over the graph filter structure to perform intersections between the
outgoing edges of two vertices. We note that in our implementation, we
perform intersection sequentially, which theoretically guarantees a
depth of $O(\sqrt{m})$. However, the parallelism of this algorithm is
$O(m^{3/2}/m^{1/2}) = O(m)$ which in practice is sufficiently high
that reducing the depth using a parallel intersection routine is
unnecessary, and can hurt performance due to using a more complicated
intersection algorithm.

Theoretically, the bounds for Triangle Counting in
Table~\ref{table:introtable} can be obtained by using a parallel
intersection method on the filter structure. The implementation of
this idea for the blocked adjacency lists in the filter structure is
identical to the $O(\log n)$ depth intersection method defined on
compressed graphs in~\cite{dhulipala18scalable}. The idea is similar
to the classic parallel intersection, or merge algorithms on two
sorted arrays~\cite{JaJa92, blelloch18notes}.

\subsection{PageRank}\label{sec:eigenvectorproblems}
\begin{benchmark}{PageRank}
Input: $G=(V, E)$, an undirected graph.\\
Output: $\mathcal{P}$, a mapping from each vertex to its PageRank
value after a single iteration of PageRank.
\end{benchmark}
We consider the PageRank algorithm, designed to rank the importance of
vertices in a graph~\cite{brin1998pagerank}. Our PageRank
implementation is described in Section~\ref{sec:eigenvector}.

%% file: inputs/experiments-apx.tex
\section{Experiments}\label{apx:experiments}

\subsection{Graph Filtering and Triangle Counting}\label{apx:filtering}
We consider the effect that the underlying block size used in the
graph filter has on the performance of our triangle counting
implementation. Recall that for our compressed inputs,
extracting a given edge from a block requires possibly (sequentially)
decompressing the full block to retrieve the desired edge. We refer to
the work done to decompress edges within the decoded blocks as the
\emph{total work} done by our triangle counting implementation.
We note that this measure is a function of both the block size used in
the graph filter, as well as the underlying graph ordering.  We refer
to the work that has to be done to perform all directed intersections
as the \emph{intersection work} done by the triangle counting
implementation, not including decoding blocks. Note that the intersection work for a graph is a fixed
quantity for a given total ordering of the vertices. The intersection
work is a lower bound on the total amount of work that has to be done,
but the total amount of work can be much larger if the blocks
containing active edges (edges in the directed graph) are spares
blocks (e.g., if each block only contains a single active edge).

\begin{table}[!t]
\centering
\centering
\begin{tabular}[!t]{l|c|c|c|c}
%\toprule
\multicolumn{1}{c|}{Graph}     &  Block Size & Intersection Work & Total Work & Time (s) \\
\midrule
 ClueWeb  &  64   & 2.24$\cdot 10^{10}$ & 7.16$\cdot 10^{10}$ & 489 \\
 ClueWeb  &  128  & 2.24$\cdot 10^{10}$  & 9.54$\cdot 10^{10}$ & 567 \\
 ClueWeb  &  256  & 2.24$\cdot 10^{10}$ & 12.8$\cdot 10^{10}$ & 683 \\
 \midrule
 Hyperlink2014 & 64 &4.8$\cdot 10^{12}$  & 101$\cdot 10^{12}$  & 5722 \\
 Hyperlink2012  & 64 &10.4$\cdot 10^{12}$ & 67.2$\cdot 10^{12}$ & 3529 \\
%\bottomrule
\end{tabular}
  \captionof{table}{\small Tradeoff between \filterbs{}, the graph
  filter block size, the number of directed wedges that must be
  checked for this graph (Intersection Work), the total amount of work
  performed decoding edges within blocks in triangle counting
  (Total Work), and the running time of triangle counting for the
  ClueWeb graph. The second half of the table reports the same
  statistics for the Hyperlink2014 and Hyperlink2012 graphs for the
  configurations in which our reported running times in
  Figure~\ref{fig:speedup_large} were run.}
\label{table:bsworktradeoff}
\end{table}

Table~\ref{table:bsworktradeoff} reports numbers displaying this
tradeoff between the \filterbs{}, the graph filter block size, total
amount of work required to decode all edges within blocks, and the
running time of the implementation. We observe that using a smaller
\filterbs{} results in a more work-efficient implementation, which
directly translates to a faster running time. We chose to use a block
size of 64 since it provided the best tradeoff between running time on
other problems considered in this paper, the size of the compressed
graph, and running time for triangle counting.

Finally, Table~\ref{table:bsworktradeoff} also reports the block
sizes, intersection work and total work required to run triangle
counting on the Hyperlink2014 and Hyperlink2012 graph with
$\filterbs{}=64$. Recall that the intersection work is a lower bound
on the total amount of work that is done, and that the total work is
the amount of decoding operations actually performed by our
implementation. We note that triangle counting on the Hyperlink2014
graph incurs a much larger penalty in terms of total work than
Hyperlink2012 (1.5x larger). The running time of the code on the
Hyperlink2014 graph is also 1.6x larger than the time for the code on
Hyperlink2012, which supports the hypothesis that the total work is
largely responsible for the running time of the triangle counting
intersection.

\subsection{Memory Usage of \emapchunk{}}\label{apx:emapchunk_mem}
Next, we study the memory improvements in terms of total memory
used by the program due to the \emapchunk{} optimization. We
report both the total memory used, and the running times for running
BFS from a given source vertex within the massive connected component
for the ClueWeb and Hyperlink2012 graphs. We compare our
implementation to the BFS implementation from GBBS, which uses the
\emapblock{} algorithm that allocates an intermediate array with
size proportional to the number of out-neighbors of a frontier.  Both
implementations use the same source vertex.

Table~\ref{table:emapchunkmemory} shows the results of the
experiments. We observe that our new chunk-based implementation
requires roughly the same order of magnitude of memory as the previous
\emapblock{} implementation, and also results in similar running
times. Both \emapblock{} and \emapchunk{} use significantly less
memory than using \emapsparse{}. Specifically, \emapchunk{} uses 1.24x
less memory than \emapsparse{} on ClueWeb, 1.27x less memory on
Hyperlink2014, and 1.31x less memory on Hyperlink2012. The memory
usage between \emapchunk{} and \emapblock{} is likely similar since
\emapblock{} only performs DRAM writes proportional to the output size
of \emap{}, which is the same number of writes (and therefore
physical pages) as performed by \emapchunk{}.

In addition, using our \emapchunk{} implementation we can run a
``sparse-only" \emap{} with limited memory. Although running only the
sparse version of a breadth-first search is not particularly
practical, it is useful for understanding the performance improvement
gained by using Beamer's direction-optimization on very large
real-world graphs. Running a similar sparse-only breadth-first search
using either \emapsparse{} or \emapblock{} causes a segmentation fault
on the Hyperlink2012 graph. The fault occurs because the algorithm
tries to allocate an 492GB array from DRAM, which is larger than the
DRAM size of our machine.
In contrast, our \emapchunk{}-based implementation successfully runs
in 38.8s and has a peak-memory usage of 120GB of memory, which is less
than $1/3$ of our machine's memory DRAM capacity. Our implementation
lets us see that using direction-optimization results in a 3.1x
speedup over running the sparse-only code for this graph.

\begin{table}[!t]
\centering
\centering
\begin{tabular}[!t]{l|c|c|c}
\multicolumn{1}{c|}{Graph} & \multicolumn{1}{c|}{Algorithm} & \multicolumn{1}{c|}{DRAM Usage} & Time (s) \\
\midrule
ClueWeb        & \emapsparse{} & 33.0GB & 2.52 \\
ClueWeb        & \emapblock{}  & 28.2GB & 2.43 \\
ClueWeb        & \emapchunk{}  & 26.6GB & 2.35 \\
\midrule
Hyperlink2014  & \emapsparse{} & 54.0GB & 5.61 \\
Hyperlink2014  & \emapblock{}  & 42.7GB & 5.20 \\
Hyperlink2014  & \emapchunk{}  & 42.3GB & 5.10 \\
\midrule
Hyperlink2012  & \emapsparse{} & 115GB  & 12.5 \\
Hyperlink2012  & \emapblock{}  & 90.3GB & 12.3 \\
Hyperlink2012  & \emapchunk{}  & 87.5GB & 12.2 \\
%\bottomrule
\end{tabular}
  \captionof{table}{\small Effect of the sparse traversal algorithm
  (\emapsparse{}, \emapblock{}, and \emapchunk{}) on the total DRAM
  usage and the running time of our BFS implementation.
} \label{table:emapchunkmemory}
\end{table}